# First-Principles Prediction of Phononic Thermal Conductivity of Silicene: a Comparison with Graphene

Xiaokun Gu and Ronggui Yang[*]

Department of Mechanical Engineering

University of Colorado at Boulder, Colorado 80309, USA

**Abstract**

There has been great interest in two-dimensional materials, beyond graphene, for both fundamental sciences and technological applications. Silicene, a silicon counterpart of graphene, has been shown to possess some better electronic properties than graphene. However, its thermal transport properties have not been fully studied. In this paper, we apply the first-principles-based phonon Boltzmann transport equation to investigate the thermal conductivity of silicene as well as the phonon scattering mechanisms. Although both graphene and silicene are two-dimensional crystals with similar crystal structure, we find that phonon transport in silicene is quite different from that in graphene. The thermal conductivity of silicene shows a logarithmic increase with respect to the sample size due to the small scattering rates of acoustic in-plane phonon modes, while that of graphene is finite. Detailed analysis of phonon scattering channels shows that the linear dispersion of the acoustic out-of-plane (ZA) phonon modes, which is induced by the buckled structure, makes the long-wavelength longitudinal acoustic (LA) phonon modes in silicene not as efficiently scattered as that in graphene. Compared with graphene, where most of the heat is carried by the acoustic out-of-plane (ZA) phonon modes, the ZA phonon modes in silicene only have ~10% contribution to the total thermal conductivity, which can also be attributed to the buckled structure. This systematic comparison of phonon transport and thermal conductivity of silicene and graphene using the first-principle-based calculations shed some light on other two-dimensional materials, such as two-dimensional transition metal dichalcogenides.

[*] Email: Ronggui.Yang@Colorado.Edu

# I. Introduction

Since its discovery, graphene has attracted great attention due to its superb material properties for both fundamental science and various technological applications.[1-3] Both theoretical and experimental work reported a very high thermal conductivity for graphene, in the range of 2000 to 5000 W/mK at room temperature.[4-6] There have been thus a significant number of studies on the mechanism of phonon transport in graphene, although consensus has not yet been reached,[7-12] and on the potential applications of graphene as an enabling thermal management material beyond its electronic and optoelectronic applications.[13-17]

Stimulated by the advances in graphene, more recently there has been great interest in many other two-dimensional materials,[18, 19] such as silicene and $MoS_2$,[20-23] which are expected to possess some of graphene's excellent properties along with other unique properties that graphene does not have. One of the most promising materials is silicene, which is a two-dimensional graphene-like honeycomb crystal made of silicon element that is expected to be more easily integrated with silicon-based semiconductor devices. Due to the similarity of the lattice structures of graphene and silicene, silicene shares many similar electronic properties with graphene. For example, the charge carrier of silicene is massless fermion just like as in graphene.[22, 24] However, compared with the planar structure of graphene, the honeycomb lattice of silicene is slightly buckled, which leads to some new characteristics. For instance, the buckled structure breaks the symmetry of the crystal, making it possible to open a bandgap by applying an electric field,[25, 26] which is a nontrivial challenge for graphene.

Unlike electronic properties, the thermal transport properties of silicene are still not well studied, though they are crucial to the reliability and performance of potential electronic and optoelectric devices that integrate silicene as a component. Due to the similarities and differences of lattice structure between silicene and graphene, it is of great interest to explore and compare the phonon transport mechanisms in these two 2-D materials, which might shed some light on the phonon transport mechanisms of other 2-D material systems which might exhibit a large variation of thermal conductivity. The presence of flexural phonons, corresponding to out-of-plane atomic vibrations, is a key signature of two-dimensional materials. In graphene, such flexural phonon modes contribute more than 50% of its large thermal conductivity.[9, 10] Until now, it has been very unclear whether the flexural phonon modes in silicene are as important as they are in graphene. For example, recent classical molecular dynamics simulations gave the conflicting conclusions depending on the empirical potentials used by different authors.[27, 28] Another important and unsettled issue in two-dimensional materials is the length dependence of the thermal conductivity. Quite a few numerical calculations have been done, but the results have been contradictory. For example, most molecular dynamics calculations, which are limited by the finite size of simulation domains, have suggested that the thermal conductivity is finite when the length exceeds a critical value,[8, 11] while some lattice dynamics calculations showed that it might be divergent.[12] On the experimental side, very recent measurements showed that the thermal conductivity of graphene keeps increasing even when the size of the graphene sheet is larger than 10 μm.[29] While most of these studies were focused on graphene, other two-dimensional materials, such as silicene, have not been well studied.

Most theoretical studies on phonon transport in silicene are based on classical molecular dynamics simulations, which heavily rely on the empirical interatomic potentials. The reported values of the thermal conductivity of silicene from these molecular dynamics simulations range about one order of magnitude, from 5 W/mK to 50 W/mK. [27, 28, 30, 31] Although some of the empirical potentials, such as the Tersoff potential[28] and the modified Stillinger-Weber potential,[27] are able to reasonably reproduce the phonon dispersion calculated from first-principles calculations, the predicted thermal conductivities of silicene are quite inconsistent. Although there have been some efforts to correctly reproduce the buckled structure of silicene,[27] the anharmonic interactions in such buckled structures are usually not taken into account when developing the empirical potentials. A more detailed investigation beyond using empirical potentials is very much desirable.

Recently, the phonon Boltzmann transport equation (PBTE) formalism with both harmonic second-order and anharmonic higher-order force constants from the first-principles calculations has been used to study the phonon transport in a wide range of semiconductors[32-35] and thermoelectric materials,[36-38] both in crystalline form and nanostructured materials (including superlattice[39, 40] and nanowires),[41, 42] and two-dimensional materials, such as graphene.[9, 43] With the accurate and reliable inputs from first-principles calculations, the calculated thermal conductivity agrees quite well with experimental measurements. Furthermore, such newly developed simulation tool can accurately predict phonon dispersion, and the phonon scattering time of each phonon mode, which is essential to understand the thermal conduction mechanisms in unknown materials.

In this paper, we study phonon transport in silicene and compare it with phonon transport in graphene. The PBTE formalism with interatomic force constants from the first-principles calculations is employed to predict the phononic thermal conductivity of silicene. The roles of different phonon branches are carefully examined. In addition, the length-dependent thermal conductivity and the corresponding mechanism are explored.

## II. Theoretical Method

Both graphene and silicene have the honeycomb structure with two bases in the primitive cell. The crystal structure and the corresponding first Brillouin zone are shown in Fig.1 (a) and (b), where the two-dimensional material are placed in the x-y plane. Since thermal transport in graphene and silicene is isotropic due to the crystal symmetry, only the thermal conductivity in the x direction is considered. Figure 1(c) shows that the two-dimensional sheet with an infinite lateral length is sandwiched between two reservoirs with a distance $L$ apart. When the two reservoirs are kept as the same temperature $T$, the local heat flux, $J$, is zero. If a small temperature difference $\Delta T$ is applied to the two reservoirs, a steady state temperature profile will be established along the x direction leading to a nonzero heat flux $J$, which can be computed by summing up the contributions from all phonon modes through

$$J = \frac{1}{(2\pi)^3} \sum_s \int \hbar \omega_{\mathbf{q}s} v^x_{\mathbf{q}s} n_{\mathbf{q}s} d\mathbf{q}, \tag{1}$$

where $\mathbf{q}s$ stands for the s-th phonon mode at $\mathbf{q}$ in the first Brillouin zone, $\hbar$ is the Planck constant, and $\omega_{\mathbf{q}s}$, $v^x_{\mathbf{q}s}$ and $n_{\mathbf{q}s}$ are the phonon frequency, group velocity in the x direction and the non-equilibrium phonon distribution function of mode $\mathbf{q}s$, respectively. After $J$ is calculated from the contributions of each phonon mode, the macroscopic thermal conductivity can then be calculated from the Fourier's law of heat conduction, $K_{xx} = J/(\Delta T/L)$.

In order to evaluate the heat flux driven by the steady-state small temperature difference, we need to calculate the phonon frequency, group velocity and phonon distribution of each phonon mode in Eq. (1), as presented below using first-principles-based phonon Boltzmann transport equation.

### A. First-principles calculations for phonon dispersion and phonon scattering rate

First-principles calculations are employed to calculate phonon dispersion and phonon scattering rate by calculating the harmonic second-order force constants and anharmonic third-order force constants which are the second and third derivatives of the total energy of the crystal with respect to the displacements of individual atoms in the crystal, respectively. The standard small displacement method, as described in ref [44] is applied. One carbon (silicon) atom noted as $\mathbf{R}\tau$, i.e., the $\tau$th basis atom in the unit cell which is represented by the translation vectors $\mathbf{R}$, in a graphene (silicene) supercell is displaced with a small distance $\Delta = 0.015$ Å from its equilibrium position along the $\pm x$, $\pm y$ and $\pm z$ directions, while other atoms remain in their equilibrium positions. The force acting on each atom due to this small displacement is recorded from the output of the first-principles calculations. The harmonic force constants with respect to the displacement of atom $\mathbf{R}\tau$ in the $\alpha$ direction and the displacement of atom $\mathbf{R}'\tau'$ in the $\beta$ direction can then be calculated as

$$\phi_{\mathbf{R}\tau,\mathbf{R}'\tau'}^{\alpha\beta} = -\frac{F_{\mathbf{R}'\tau'}^{\beta}\left(u_{\mathbf{R}\tau}^{\alpha} = \Delta\right) - F_{\mathbf{R}'\tau'}^{\beta}\left(u_{\mathbf{R}\tau}^{\alpha} = -\Delta\right)}{2\Delta} \quad (2)$$

where $F$ is the force and $u$ is the displacement. Taking the advantage of the periodicity and symmetry of the lattice structure, moving one atom is enough to extract all independent harmonic force constants. With the harmonic force constants calculated from the first-principle calculations, the dynamical matrix $D$ with the pairs $(\tau,\alpha)$ and $(\tau',\beta)$ as indices can then be solved for phonon dispersion,

$$D_{\tau\tau'}^{\alpha\beta}(\mathbf{q}) = \frac{1}{\sqrt{M_\tau M_{\tau'}}} \sum_{\mathbf{R}'} \phi_{0\tau,\mathbf{R}'\tau'}^{\alpha\beta} e^{i\mathbf{q}\cdot\mathbf{R}'}, \quad (3)$$

where $M_\tau$ is the atomic mass of the $\tau$th basis of the primitive cell. The phonon frequency $\omega_{qs}$ is the square root of the $s$-th eigenvalue of the dynamical matrix and the group velocity $v_{qs}^x$ is calculated as $\partial \omega_{\mathbf{q}s} / \partial q_x$. The third-order force constants are calculated by the similar procedures but moving two atoms simultaneously. The third-order force constants with respect to the displacements of atom $\mathbf{R}\tau$, atom $\mathbf{R}'\tau'$ and atom $\mathbf{R}''\tau''$ along the $\alpha$, $\beta$, and $\gamma$ directions are written as

$$\psi^{\alpha\beta\gamma}_{\mathbf{R}\tau,\mathbf{R}'\tau',\mathbf{R}''\tau''} = -\frac{1}{4\Delta^2}\left[F^{\gamma}_{\mathbf{R}''\tau''}\left(u^{\alpha}_{\mathbf{R}\tau}=\Delta, u^{\beta}_{\mathbf{R}'\tau'}=\Delta\right) - F^{\gamma}_{\mathbf{R}''\tau''}\left(u^{\alpha}_{\mathbf{R}\tau}=\Delta, u^{\beta}_{\mathbf{R}'\tau'}=-\Delta\right)\right.$$
$$\left. -F^{\gamma}_{\mathbf{R}''\tau''}\left(u^{\alpha}_{\mathbf{R}\tau}=-\Delta, u^{\beta}_{\mathbf{R}'\tau'}=\Delta\right) + F^{\gamma}_{\mathbf{R}''\tau''}\left(u^{\alpha}_{\mathbf{R}\tau}=-\Delta, u^{\beta}_{\mathbf{R}'\tau'}=-\Delta\right)\right]. \quad (4)$$

With the third-order force constants calculated from the first-principles, the three-phonon scattering rate can be calculated using the Fermi's golden rule. Once three phonons satisfy the momentum conservation condition $\mathbf{q} \pm \mathbf{q}' = \mathbf{q}'' + \mathbf{G}$, with $\mathbf{G}$ representing a reciprocal vector, the transition probabilities of the three-phonon processes $\mathbf{q}s + \mathbf{q}'s' \rightarrow \mathbf{q}''s''$ and $\mathbf{q}s \rightarrow \mathbf{q}'s' + \mathbf{q}''s''$ are written as[45]

$$W^{\mathbf{q}''s''}_{\mathbf{q}s,\mathbf{q}'s'} = 2\pi n_{\mathbf{q}s} n_{\mathbf{q}'s'} (n_{\mathbf{q}''s''}+1)|V_3(-\mathbf{q}s,-\mathbf{q}'s',\mathbf{q}''s'')|^2 \delta(\omega_{\mathbf{q}s}+\omega_{\mathbf{q}'s'}-\omega_{\mathbf{q}''s''})$$

$$W^{\mathbf{q}'s',\mathbf{q}''s''}_{\mathbf{q}s} = 2\pi n_{\mathbf{q}s} (n_{\mathbf{q}'s'}+1)(n_{\mathbf{q}''s''}+1)|V_3(-\mathbf{q}s,\mathbf{q}'s',\mathbf{q}''s'')|^2 \delta(\omega_{\mathbf{q}s}-\omega_{\mathbf{q}'s'}-\omega_{\mathbf{q}''s''}), \quad (5)$$

where the delta function denotes the energy conservation condition $\omega_{\mathbf{q}s} \pm \omega_{\mathbf{q}'s'} = \omega_{\mathbf{q}''s''}$ for the three-phonon scattering process, the + and - signs represent the annihilation and decay processes, respectively and $V_3$ is the three-phonon scattering matrix

$$V_3(\mathbf{q}s,\mathbf{q}'s',\mathbf{q}''s'') = \left(\frac{\hbar}{8N_0 \omega_{\mathbf{q}s}\omega_{\mathbf{q}'s'}\omega_{\mathbf{q}''s''}}\right)^{1/2} \sum_{\tau}\sum_{\mathbf{R}'\tau'}\sum_{\mathbf{R}''\tau''}\sum_{\alpha\beta\gamma} \psi^{\alpha\beta\gamma}_{0\tau,\mathbf{R}'\tau',\mathbf{R}''\tau''} e^{i\mathbf{q}'\cdot\mathbf{R}'} e^{i\mathbf{q}''\cdot\mathbf{R}''} \frac{e^{\tau\alpha}_{\mathbf{q}s} e^{\tau'\beta}_{\mathbf{q}'s'} e^{\tau''\gamma}_{\mathbf{q}''s''}}{\sqrt{M_\tau M_{\tau'} M_{\tau''}}}.$$
(6)

where $e$ is the eigenvector of the dynamical matrix, $N_0$ is the number of unit cells. We have considered both the normal (when $\mathbf{G}=\mathbf{0}$) and the Umklapp (when G represents a reciprocal vector) three-phonon processes in current PBTE framework. All the first-principles calculations are performed with the first-principle software package Quantum-Espresso[46] using norm-conserving pseudopotentials. Plane-wave basis sets with kinetic-energy cut-offs of 80 Ry and 60 Ry are used in the graphene and silicene calculations, respectively. A 28×28×1 Monkhorst-Pack *k*-point mesh is used to relax the structure. The kinetic energy cut-off and *k*-point mesh are carefully tested so that the calculated thermal conductivity will not change even if the number of *k*-point is decreased to the half and the kinetic-energy cut-offs are reduced by 20 Ry. The cutoffs of interaction are applied to the atoms within the ninth and third nearest neighbors for harmonic and third-order force constants, respectively. The dispersion relation is found not to change by taking more neighbors into account. The translational invariance is imposed to make the set of interatomic force constants physically correct. Because we extract the interatomic force constants when atoms are in their equilibrium position, the effect of temperature on the interatomic force constants are not taken into account. For most materials, the force constants at 0K provide reasonable description of the phonon transport processes.[47]

### B. Numerical solution of phonon Boltzmann transport equation

The linearized PBTE is then solved to find the phonon distribution function $n_{\mathbf{q}s}$ in Eq. (1) considering the balance between phonon diffusion driven by the small temperature difference and phonon scatterings due to various scattering mechanisms.

In addition to the intrinsic three-phonon scattering described in Section II.A, the boundary scattering plays an important role at low temperature and in the low-dimensional systems. The relaxation time of boundary scattering of phonons for finite size sample can be modeled using the partially specular and partially diffusive model, written as[48]

$$\tau_{\mathbf{q}s}^{\mathrm{B}} = \frac{1+p}{1-p}\frac{L}{2|v_{\mathbf{q}s}^{x}|}, \qquad (7)$$

where $p$ and $1-p$ are the fractions of phonons that are specularly and diffusively scattered at the interface, respectively. By comparing the wavelength of the dominant phonon modes from 200 K to 800 K and the roughness of the boundary, we expect that the boundary scattering is predominantly diffusive.[49] Considering that directly quantify the specularity $p$ is still a challenge, we use the fully diffusive scattering, i.e. $p=0$, to model the boundary scattering. The thermal conductivity for a partially specular and partially diffuse boundary scattering can be extrapolated without significant difficulty. Such treatment of boundary scattering has been proved to correctly reproduce the transition between diffusive transport and the ballistic transport and has been employed in the studies on the phonon transport in many other low-dimensional materials, such as carbon nanotubes[50] and graphene.[10]

Taking both the intrinsic three-phonon scatterings and the phonon-boundary scatterings into account, the linearized PBTE is now expressed as[45]

$$v_{\mathbf{q}s}^{x}\frac{\partial T}{\partial x}\frac{\partial n_{\mathbf{q}s}^{0}}{\partial T} = \sum_{\mathbf{q}'s',\mathbf{q}''s''}\left[\left(W_{\mathbf{q}''s''}^{\mathbf{q}s,\mathbf{q}'s'} - W_{\mathbf{q}s,\mathbf{q}'s'}^{\mathbf{q}''s''}\right) + \frac{1}{2}\left(W_{\mathbf{q}'s',\mathbf{q}''s''}^{\mathbf{q}s} - W_{\mathbf{q}s}^{\mathbf{q}'s',\mathbf{q}''s''}\right)\right] - \frac{n_{\mathbf{q}s} - n_{\mathbf{q}s}^{0}}{\tau_{\mathbf{q}s}^{\mathrm{B}}}. \qquad (8)$$

To solve Eq. (8), we can write $n_{\mathbf{q}s}$ as $n_{\mathbf{q}s}^{0} + n_{\mathbf{q}s}^{0}\left(n_{\mathbf{q}s}^{0}+1\right)\varphi_{\mathbf{q}s}$ with the unknown deviation function $\varphi_{\mathbf{q}s}$,[51] which is a function of the temperature gradient. When the temperature gradient vanishes, $\varphi_{\mathbf{q}s}$ would be zero and the non-equilibrium phonon distribution function recovers to the equilibrium one. Expanding $\varphi_{\mathbf{q}s}$ into the Taylor series with respect to the temperature gradient and neglecting the higher-order terms in the limit of a small temperature gradient, $\varphi_{\mathbf{q}s}$ can further be assumed proportional to the temperature gradient, $\varphi_{\mathbf{q}s} = \frac{\partial T}{\partial x}F_{\mathbf{q}s}$. We then can write $n_{\mathbf{q}s} = n_{\mathbf{q}s}^{0} + n_{\mathbf{q}s}^{0}\left(n_{\mathbf{q}s}^{0}+1\right)\frac{\partial T}{\partial x}F_{\mathbf{q}s}$.[44, 45] With this substitution, Eq. (8) is then recast to the form,[44, 45]

$$v_{\mathbf{q}s}^{x}\frac{\partial n_{\mathbf{q}s}^{0}}{\partial T}=\sum_{\mathbf{q}'s',\mathbf{q}"s"}\left[\tilde{W}_{\mathbf{q}s,\mathbf{q}'s'}^{\mathbf{q}"s"}\left(F_{\mathbf{q}"s"}-F_{\mathbf{q}'s'}-F_{\mathbf{q}s}\right)+\frac{1}{2}\tilde{W}_{\mathbf{q}s}^{\mathbf{q}'s',\mathbf{q}"s"}\left(F_{\mathbf{q}"s"}+F_{\mathbf{q}'s'}-F_{\mathbf{q}s}\right)\right]-\frac{n_{\mathbf{q}s}^{0}\left(n_{\mathbf{q}s}^{0}+1\right)F_{\mathbf{q}s}}{L/2\left|v_{\mathbf{q}s}^{x}\right|},$$

(9)

where $\tilde{W}_{\mathbf{q}s,\mathbf{q}'s'}^{\mathbf{q}"s"}$ and $\tilde{W}_{\mathbf{q}s}^{\mathbf{q}'s',\mathbf{q}"s"}$ are the equilibrium transition probabilities, sharing the same form as Eq. (5) but with the non-equilibrium phonon distribution functions ($n_{\mathbf{q}s}$, $n_{\mathbf{q}'s'}$, and $n_{\mathbf{q}"s"}$) replaced by the equilibrium functions ($n_{\mathbf{q}s}^{0}$, $n_{\mathbf{q}'s'}^{0}$, and $n_{\mathbf{q}"s"}^{0}$).

In order to perform the summation of the three-phonon scattering events in Eq. (9) to solve for the phonon distribution function $F_{\mathbf{q}s}$, the first Brillouin zone is discretized into a $N \times N$ Γ-point centered grid, as shown in Fig. 1(b). The grid points are located at $\mathbf{q}=\frac{m}{N}\mathbf{b}_{1}+\frac{n}{N}\mathbf{b}_{2}$, where $\mathbf{b}_{1}$ and $\mathbf{b}_{2}$ are reciprocal primitive lattice vectors, and $m$ and $n$ are integers. The delta function in Eq. (5) that guarantees that the energy conservation in the three-phonon processes is replaced by a Gaussian function with adaptive broadening,[41, 52]

$$\delta\left(\omega_{\mathbf{q}s}\pm\omega_{\mathbf{q}'s'}-\omega_{\mathbf{q}"s"}\right)=\frac{1}{\sqrt{\pi}\sigma}\exp\left(-\left(\omega_{\mathbf{q}s}\pm\omega_{\mathbf{q}'s'}-\omega_{\mathbf{q}"s"}\right)^{2}/\sigma^{2}\right) \quad (10)$$

where

$$\sigma=\left|\nabla\omega_{\mathbf{q}'s'}-\nabla\omega_{\mathbf{q}"s"}\right|\Delta q \quad (11)$$

and $\Delta q$ is the distance between neighboring sampling points. Contrasting with the common practice of setting the broadening parameter $\sigma$ to a constant value for all the three-phonon scattering events,[43, 53] the adaptive scheme is able to take the non-uniformity of the energy spacing into account. Moreover, it avoids the relatively arbitrary choice of $\sigma$ in the constant broadening parameter scheme.

The set of linear equations Eq. (9), with respect to $F_{\mathbf{q}s}$, can then be self-consistently solved. Here we employ the biconjugate gradient stabilized method (Bi-CGSTAB), a variant of the conjugate gradient algorithm, to iteratively solve it. The details of the algorithm are provided in ref [54]. This algorithm successfully removes the numerical instability [53, 55, 56] that appeared in the original iterative algorithm proposed by Omini and Sparavigna.[55]

After $F_{\mathbf{q}s}$ is calculated, the thermal conductivity of the two-dimensional material can be written as

$$K_{xx}(x) = \frac{2\hbar}{N_0 \sqrt{3} a_0^2 h} \sum_{\mathbf{q}s} \omega_{\mathbf{q}s} v_{\mathbf{q}s}^x n_{\mathbf{q}s}^0 \left(n_{\mathbf{q}s}^0 + 1\right) F_{\mathbf{q}s}. \tag{12}$$

Aside from the iterative solution of Eq. (9), if $F_{\mathbf{q}'s'}$ and $F_{\mathbf{q}''s''}$ are set to be zero in Eq. (10), we can easily solve $F_{\mathbf{q}s}$ and obtain the widely used solution of PBTE under single-mode relaxation time approximation (RTA). The thermal conductivity under single-mode RTA is

$$K_{xx} = \frac{2\hbar^2}{N_0 \sqrt{3} a_0^2 h k_B T^2} \sum_{\mathbf{q}s} \omega_{\mathbf{q}s}^2 \left(v_{\mathbf{q}s}^x\right)^2 n_{\mathbf{q}s}^0 \left(n_{\mathbf{q}s}^0 + 1\right) \tau_{\mathbf{q}s} \tag{13}$$

where $\tau_{\mathbf{q}s} = 1/(1/\tau_{\mathbf{q}s}^{ph} + 1/\tau_{\mathbf{q}s}^B)$ is the relaxation time of mode $\mathbf{q}s$ with the relaxation time due to phonon-phonon scattering, $\tau_{\mathbf{q}s}^{ph}$, which is expressed as

$$\tau_{\mathbf{q}s}^{ph} = \frac{n_{\mathbf{q}s}^0 (n_{\mathbf{q}s}^0 + 1)}{\sum_{\mathbf{q}'s',\mathbf{q}''s''} \left( \tilde{W}_{\mathbf{q}s,\mathbf{q}'s'}^{\mathbf{q}''s''} + \frac{1}{2} \tilde{W}_{\mathbf{q}s}^{\mathbf{q}'s',\mathbf{q}''s''} \right)}. \tag{14}$$

### III. Results and Discussion

In Sec. III(A) and (B), we first compare the phonon dispersion and report the intrinsic phononic thermal conductivity of graphene and silicene. In Sec. III(C), the role of the flexural modes on the phonon transport in both graphene and silicene is discussed. Finally, the length-dependent thermal conductivity is analyzed by exploring the phonon scattering mechanism in Sec. III(D).

### A. Phonon dispersion

Our DFT calculation yields an equilibrium lattice constant $a_0$ = 3.824 Å and buckling distance $\delta$ = 0.42 Å for silicene. These lattice parameters are in good agreement with previous studies.[57] For graphene, a lattice constant of $a_0$ = 2.437 Å is obtained after relaxation. We use the nominal layer thicknesses $h$ = 0.335 nm and $h$ = 0.42 nm for graphene and silicene, corresponding to the van der Waals radii of carbon and silicon atoms[28] to report our results.

Figure 2 shows the phonon dispersion of graphene and silicene in the high-symmetry directions. Due to two bases in the primitive unit cell of graphene and silicene, there are six phonon branches for each material: longitudinal acoustic (LA), transverse acoustic (TA), flexural out-of-plane acoustic branches (ZA), longitudinal optical (LO), transverse optical (TO), and flexural out-of-plane optical (ZO) branches. The frequency of the LO (or TO) mode of graphene at the Γ point is 1630 cm$^{-1}$, which is in good agreement with the measured G peak Raman signal.[58] The calculated phonon dispersion of silicene is identical to other calculations considering different ranges of interaction cutoff and using different source codes in literature.[25, 59, 60] In the phonon dispersion of graphene and silicene shown in Fig. 2, we notice that the ZA branch near the Γ point shows a quadratic trend for both materials, which is a typical features of two-dimensional materials and could be explained by the macroscopic elastic theory of thin plates.[1, 61] However, while the phonon dispersion of graphene's ZA branch follows exactly the $\omega_{ZA} \propto q^2$ relation near the Γ point, a small but non-zero sound velocity for the ZA branch of silicene with a value of 1010 m/s is observed, which is an order of magnitude smaller than that of LA and TA branches. This observation is consistent with density functional perturbation theory calculation from other studies.[62] The different shape of the ZA phonon dispersion makes the group velocity and equilibrium phonon distribution function different for graphene and silicene near the zone center: $v \propto q$ and $n^0 \propto q^{-2}$ for graphene while $v \propto q^0$ and $n^0 \propto q^{-1}$ for silicene. This difference on the shape of ZA branches near the zone center comes from the structural difference of the two materials: graphene is planar while silicene is a buckled structure. From a microscopic point of view, the $\omega_{ZA} \propto q^2$ relation near the Γ point for graphene is the direct result of two factors: 1) the rigid rotational invariance when the material rotates along any axis within the plane and 2) the decoupling of in-plane modes and out-of-plane modes due to the one-atom-thick nature of graphene. The details of the derivation can be found in the supplemental material of ref. [9]. However, silicene does not obey the rigid rotational invariance nor does it exhibit the decoupling of in-plane and out-of-plane phonon modes due to its buckled structure. In Appendix A1, we provide a proof to show that the dispersion of the ZA branch of silicene near the Gamma point always contains a linear component, just as it does in an acoustic phonon branch in a conventional three-dimensional material.

It is worthwhile to mention that the quadratic dispersion of the ZA branch might lead to problematically strong scattering for in-plane acoustic phonons.[9, 56] Mariani and von Oppen[63] theoretically proved that the coupling of bending and stretching modes renormalizes the quadratic ZA dispersion to $\omega \propto q^{3/2}$ as $q \to 0$. In this work, the dispersion relation for ZA branch near the zone center in graphene is slightly modified to make the ZA dispersion follow the derived renormalized dispersion relation.[63] More discussions are given in Appendix B. In Sec. III(D), we show that we can get rid of the problematically strong scattering using the dispersion relation $\omega \propto q^{3/2}$ for ZA branch.

### B. Intrinsic thermal conductivity of graphene and silicene

Figure 3 shows the dependence of calculated thermal conductivity of graphene and silicene with the sample size $L = 3\mu$m as a function of the number of $q$-grids used. Thermal conductivities are found to increase with the number of $q$-grid sampling points. Such a dependence on the $q$-grids can be qualitatively explained by the single-mode RTA. The thermal conductivity formula under the single-mode RTA can be expressed as

$$K \propto \sum_s \int \left(v_{\mathbf{q}s}^x\right)^2 \omega_{\mathbf{q}s}^2 n_{\mathbf{q}s}(n_{\mathbf{q}s}+1)\tau_{\mathbf{q}s} d\mathbf{q} \tag{15}$$

If a coarse $q$-grid is used, the zone center will not be well sampled. Taking advantage of the isotropicity of the frequencies near the zone center, the double integrals over $\mathbf{q}$ vector in Eq. (15) can be transformed to a one-dimensional integral over scale $q$. The difference between the thermal conductivities calculated from an idealized infinitely dense grid and that from a finite $q$-grid can be estimated as

$$\Delta K \propto \sum_s \int_0^{q_{\text{cut}}} v_{\mathbf{q}s}^2 \omega_{\mathbf{q}s}^2 n_{\mathbf{q}s}(n_{\mathbf{q}s}+1)\tau_{\mathbf{q}s} q dq, \tag{16}$$

where $q_{\text{cut}}$ is on the order of the difference between two neighboring $q$ points. Eq. (16) also gives the estimation of the contribution from long-wavelength phonons to the total thermal conductivity. If the integrand scales with $q^n$ with $n > 0$ as $q \to 0$, the error $\Delta K$ converges rapidly to zero as $q_{\text{cut}} \to 0$. Otherwise, the $\Gamma$ point becomes a singular point, which makes $\Delta K$ highly dependent on $q_{\text{cut}}$ and causes the $q$-grid dependence on the calculated thermal conductivity as shown in Fig. 3. Since the long-wavelength acoustic phonons tend to be less scattered by other phonons, the relaxation time due to phonon scattering approaches infinity as $q \to 0$. The relaxation time of the boundary scattering $\tau_{\mathbf{q}s} = L/2|v_{\mathbf{q}s}|$ plays a more significant role than the intrinsic phonon-phonon scattering. As a result, $\Delta K$ scale as $q_{cut}^2$ for acoustic phonons with linear dispersion relation.

In order to get rid of the dependence of the thermal conductivity on the $q$-grid, we calculate the thermal conductivity using several $q$-grids with different number of sampling points up to 16384 ($N = 128$) and then extrapolate the thermal conductivity to an infinitely dense grid using the relation $\Delta K \propto q_{cut}^2$, based on the above analysis, as shown in Fig. 3. We also note that $\Delta K$ might not be exactly proportional to $q^2$ since the effective phonon relaxation time also has the contributions from the intrinsic phonon-phonon scattering. To test the robustness of the extrapolation process, we also perform linear and cubic extrapolations. The difference between different extrapolation methods is within 2% of the thermal conductivity.

To validate the approach we employ, the calculated thermal conductivity of graphene was compared with the data available from literature, including both numerical simulations and experimental measurement. Figure 4 shows the thermal conductivity of graphene as a function of temperature. The solid, dashed, dashed-dotted black lines are our calculated thermal conductivity for graphene with $L$=100 μm, 10 μm and 3 μm using the modified ZA dispersion as discussed in Appendix B. For completeness, we also show the thermal conductivity of the sample with $L$=100 μm using the ZA dispersion from DFT calculations, as indicated by the blue dotted line. It is clear that the modified ZA dispersion only slightly changes the value of the thermal conductivity of graphene. The green line is theoretical predictions from other groups' work using similar PBTE formalism.[7] Our calculated values for a graphene sheet with $L$=100 μm is very close to Singh *et al*'s calculation for an infinitely large graphene sheet,[7] where they employed the optimized Tersoff potential[64] to describe the interatomic interaction of graphene. The slightly larger value obtained in our prediction is probably due to the different force constants used. Figure 4 also includes several experimental measurements using Raman spectroscopy.[65-68] Since in thermal conductivity measurement experiments graphene sheets are always suspended above holes with diameters of several microns, it is more reasonable to compare these experimental results to our calculated values with $L=$ 3 μm to 10 μm. Although the measurement data is quite scattered due to experimental accuracy and sample processing, the values from our theoretical calculations follow a similar trend as these experimental measurements. The results shown in Fig. 4 indicate the PBTE approach with first-principles force constants is able to give reliable predictions for the thermal conductivity of two-dimensional materials.

Figure 5 shows the temperature dependence of the thermal conductivity of silicene for different sample sizes. Dramatically different from graphene whose thermal conductivity is comparable to or even larger than its bulk forms, graphite or diamond, the thermal conductivity of silicene is an order of magnitude lower than that of bulk silicon. For example, it is around 26 W/mK for a silicene sheet with $L=10\mu m$ at 300 K while the thermal conductivity of silicon is ~140 W/mK at room temperature. Indeed the thermal conductivity of silicene with $L=10\mu m$ is comparable to the conductivity of silicon nanowires with a diameter of 55 nm[69] and thin films with a thickness of 20 nm.[70] When sample length decreases, the thermal conductivity is further reduced and shows weaker temperature dependence due to the stronger boundary scattering.

We notice that a very recent study using molecular dynamics simulations reported a much smaller thermal conductivity for silicene, ~5 W/mK, than our results. In their work, equilibrium molecular dynamics simulations is used to study the thermal conductivity of silicene with a modified Stillinger-Weber potential,[27] which was originally developed for bulk silicon. Although the empirical potential they used has been optimized to better match the phonon dispersion of the silicene from DFT calculations than the original Stillinger-Weber potential,[71] the difference likely originates from two reasons: 1). the phonon dispersion relations from the empirical potential, especially the acoustic branches, are still different from those from DFT calculations; 2). anharmonic properties which are essential for predicting phonon-phonon scattering rate were not taken into account in the development of the potential. In addition to the quantitative difference on the predicted thermal conductivity, first-principle-based PBTE method provides much more detailed information of each phonon mode as discussed in this work.

To understand why silicene has a much lower thermal conductivity than silicon and graphene, we examine the contributions of the different phonon branches in graphene and silicene, as shown in Fig. 6. When $L=$ 10 μm, while the ZA branch contributes about 75% of the large thermal conductivity of graphene, only around 7.5% is conducted by the ZA branch in silicene. The LA and TA branches together contribute 20% and 70% to the total thermal conductivity for graphene and silicene, respectively. From this comparison, one can conclude that the overall low thermal conductivity of silicene should be related to the strong scattering of the ZA modes. It is noted that the relative contribution of the ZA modes to the thermal conductivity of graphene is different from some other theoretical works, including the calculations using relaxation time approximation[73, 74] and MD simulations.[75, 76] Considering that the approximate nature of the empirical expressions for phonon relaxation time or empirical potentials employed, these methods might not give accurate results, though the studies could still provide valuable insights.

### C.  The role of the ZA branch

Considering that there exists the significant difference in the contribution of the ZA modes to the total thermal conductivity in graphene and silicene, it is worthwhile to perform a more detailed investigation of the scattering mechanism of the ZA modes in graphene and silicene. Figure 7 shows the calculated scattering rates of acoustic phonons due to phonon-phonon scattering. The scattering rate is defined as the inverse of the relaxation time, $\Gamma_{qs} \propto q$ characterizing the strength of the scattering mechanism. The larger the scattering rate is, the more likely a phonon is to be scattered, i.e., with shorter lifetime. Figure 7 shows clearly that the scattering rates of ZA modes in graphene modes are much smaller than the in-plane acoustic phonon modes. However, the scattering rates for the ZA modes in silicene are comparable to that of the other acoustic phonon modes.

To understand the difference in the scattering rates of ZA modes in silicene and graphene, we decompose the total scattering rates of the ZA modes into different scattering channels, as shown in Fig. 8. Figure 8(a) shows that the ZA phonons in graphene are dominantly scattered through the absorption processes in which two ZA phonons combine to produce one in-plane acoustic phonon (LA or TA). However, Figure 8(b) shows that absorption processes dominates the scattering channel of ZA modes where a ZA phonon is produced involving two in-plane LA or TA acoustic phonons. The absorption processes also contributes to ~10% of the total scattering close to the zone center. These scattering channels are not observed in graphene. This is because of a symmetry selection rule[10] in graphene: for one-atom-thick materials, reflection symmetry makes the third-order force constants involving an odd number of z components vanish. As a result, scattering with odd number of out-of-plane modes, such as ZA+TA->LA and ZA+ZA->ZA, could never happen. However, due to the buckled structure in silicene, the symmetry selection rule does not apply. Therefore, the out-of-plane ZA phonon modes in silicene have more scattering channels than that in graphene as long as there is another out-of-plane or in-plane phonon available. To further explore the importance of the scattering channels involving an odd number of out-of-plane phonons on the total scattering rate of the ZA modes, Fig. 8(b) also shows the scattering rate when these scattering channels are tuned off. The scattering rates of the ZA phonons, especially the long-wavelength phonons are greatly suppressed and reach a value comparable to the scattering rates in graphene. This investigation testifies that the scattering rate of the ZA modes in silicene is greatly increased, compared with graphene, due to the buckled structure where the symmetry selection rule that applies in graphene is broken.

### D. Length dependence of the thermal conductivity

One might conceive that the thermal conductivity of silicene would saturate quickly as the length increases because the thermal conductivity of silicene is low and the phonon mean free path in silicene might be short. However, we observe a length-dependent thermal conductivity even when the length of the silicene sample is larger than 30 μm, as discussed below.

Figure 9 shows the thermal conductivity of graphene and silicene at 300 K as a function of sample size $L$. While the thermal conductivity of graphene appears to converge when the sample size is larger than $100 \mu$m, the thermal conductivity of silicene increases logarithmically with the sample size in the range studied in this work. The thermal conductivity of silicene increases from 18 W/mK to 28 W/mK when the size is increased from 0.3 μm to 30 μm. The unbounded thermal conductivity occurs in the whole temperature range studied in this work. The divergent (unbounded) thermal transport phenomenon of silicene with the increase of sample size can also be verified by examining the dependence of the thermal conductivity on the $q$-grid when the boundary scattering is absent, as shown in the inset of Fig. 9(b). To understand such anomalous phenomenon, the contributions of thermal conductivity from different phonon branches are also shown in Fig. 9. As the sample size increases, the contributions from the optical modes and the out-of-plane ZA acoustic phonon modes of silicene become constant when $L > 1$ μm. The continuous increase of the thermal conductivity with respect to the sample size is indeed due to the in-plane acoustic phonons. Therefore, we then examine how the thermal conductivity of in-plane acoustic phonons varies with the sample size. Figure 10 shows the accumulated thermal conductivity of the LA and TA in-plane phonon modes of two silicene sheets with different sample size, $L = 10$ μm and $L = 30$ μm, as a function of phonon frequency. It is obvious that the difference of the thermal conductivity between the $L = 10$ μm and $L = 30$ μm samples is caused by the low-frequency/long-wavelength phonons.

With a known linear dispersion of long-wavelength in-plane acoustic phonons as discussed in Sec. III(A), we are able to find out whether the thermal conductivity becomes unbounded if the relaxation time is known. According to Eq. (16), a finite thermal conductivity will be obtained if the relaxation time of the long-wavelength in-plane phonons which has linear dispersion follow $q^n$ if $n > -2$. We thus perform a careful investigation on how long-wavelength in-plane acoustic phonons are scattered in graphene and silicene.

In Fig. 11, the scattering rates of the LA and TA in-plane acoustic phonons are plotted using a log-log scale. Figure 11(a) shows that the scattering rates of the in-plane acoustic phonons in graphene linearly decreases to zero with respect to the wavenumber $q$ at the zone center as $q \to 0$. We noticed that the LA and TA phonon modes of graphene in the region $0 < q < 0.05(2\pi/a_0)$ are scattered almost exclusively by decaying into two out-of-plane ZA modes. We can now follow an approach similar to that presented by Bonini et al,[9] where they studied the analytic limit as $q \to 0$ by considering the decaying process of the scattering of LA and TA modes into two ZA modes on the quadratic dispersion. In a similar analysis shown in Appendix C, we find that the analytic limit of the scattering rate is $\Gamma_{\mathbf{q}s} \propto q$ when in-plane acoustic phonon modes are scattered to two phonons on the ZA branch with $\omega \propto q^{3/2}$ dispersion, which can explain well the numerical results observed. According to $\Gamma_{\mathbf{q}s} = 1/\tau_{\mathbf{q}s} \propto q$, the thermal conductivity of the long-wavelength LA and TA branches of graphene should be finite, since the relaxation time follows the relation that ensures the finite thermal conductivity, $\tau_{\mathbf{q}s} \propto q^n$ with $n = -1 > -2$. In addition, the modified dispersion for ZA modes makes the product $\omega_{\mathbf{q}s}\tau_{\mathbf{q}s}$ approaches constant values, instead of zero as observed when the quadratic dispersion is used.[9,56] From our numerical calculation, the constant values are around 8 for LA modes and 14 for TA modes, both of which are much larger than unity. The larger-than-unity constant values of $\omega_{\mathbf{q}s}\tau_{\mathbf{q}s}$ ensure the condition for the existence of phonons as elementary excitations[72] is valid for long-wavlength in-plane LA and TA phonons of graphene. Therefore, by considering the renormalization of ZA phonons[63] as discussed in Appendix B, the problematic large scattering of the long-wavelength in-plane acoustic phonons in graphene can be avoided.

Figure 11(b) shows that the scattering rates of silicene also approach zero when the wavenumber $q$ goes to zero, but with a much steeper slope than that for graphene. It is found that the scattering of the in-plane LA and TA phonon modes also mainly comes from the decay processes into two ZA modes at the small wavenumber regime $0.01(2\pi/a_0) < q < 0.1(2\pi/a_0)$. However, the out-of-plane ZA phonon branch on a linear dispersion is not able to scatter the LA and TA phonons as efficiently as that in graphene. The scattering process for three acoustic phonons on linear dispersions are derived in detail in Appendix D which shows that the inverse of the scattering rate, or the relaxation time, scales with $q^n$ with $n=-3<-2$ as $q \to 0$.

In addition to the decay processes, a long-wavelength LA or TA mode, $\mathbf{q}s$, in silicene can also annihilate with another mode, $\mathbf{q}'s'$, and generate the third mode $\mathbf{q}''s''$ which needs to satisfy the momentum conservation $(\mathbf{q}''=\mathbf{q}+\mathbf{q}'+\mathbf{G})$. The analytical limits of the scattering rate for the annihilation processes are also derived in Appendix D. If modes $\mathbf{q}'s'$ and $\mathbf{q}''s''$ are on different branches, for example, LA+LA/TA->TA/LA, the scattering rate of the in-plane acoustic phonons scales with $q^2$ as $q \to 0$. If modes $\mathbf{q}'s'$ and $\mathbf{q}''s''$ are on the same branch, such as TA+ZO->ZO, the scattering rate of the in-plane TA branch has the form $\Gamma_{\mathbf{q}s}=1/\tau_{\mathbf{q}s} \propto q$ as $q \to 0$. This linear dependence of the scattering rate on the wavenumber $q$ for the in-plane TA modes ensures that the condition $\tau_{\mathbf{q}s} \propto q^n$ with $n>-2$ is satisfied as $q \to 0$ and thus the thermal conductivity of TA branch is finite.

However, unlike in-plane TA modes, the annihilation process for a long-wavelength in-plane LA phonon with two phonon modes on the same branch are always prohibited. This is because the group velocity of the long-wavelength in-plane LA modes, $v_{\mathrm{LA}}$, is the largest of all phonon modes so that the frequency of the long-wavelength in-plane LA mode $\mathbf{q}s$, which can be estimated by $v_{\mathrm{LA}}q$ due to the linear dispersion, is always larger than the frequency difference between the modes $\mathbf{q}'s'$ and $\mathbf{q}''s''$, which can be written as $|\mathbf{v}_{\mathbf{q}'s'} \cdot \mathbf{q}|$ $(\leq v_{\mathbf{q}'s'}q < v_{\mathrm{LA}}q)$ as $q \to 0$. As a result, the annihilation scattering with two phonon modes on the different branches becomes the dominant scattering mechanism for long-wavelength LA phonon as $q \to 0$.

The total scattering rate of the long-wavelength LA phonons would follows $q^2$ as $q \to 0$. According to Eq. (16), the thermal conductivity of the long-wavelength LA modes is $\Delta K \propto \int_0^{q_{cut}} q^{-1} dq$, leading to a divergent thermal conductivity when the boundary scattering is absent. When the boundary scattering is included, $\Delta K \propto \int_0^{q_{cut}} \frac{q}{q^2 + L} dq \propto \ln L$. This analytical limit is consistent with the observation of our numerical results, as shown in Fig. 9(b). However we wanted to that the derived scaling relation of the relaxation time or the scattering rate with respect to the wavenumber $q$ for long-wavelength in-plane phonon modes would hold in the region very close to the zone center. The logarithmic dependence on the sample length we observe should be the outcome of the combination of different phonon branches, not just from the long-wavelength LA phonons.

We found a recent work on the prediction of the thermal conductivity of silicene from relaxation time approximation with interatomic force constants from first-principles calculations.[73] In their work, only three-phonon processes are considered and the calculated thermal conductivity is about 9 W/mK for an infinitely large silicene sheet, much smaller than the results from our calculations. The relaxation times of long wave-length acoustic phonon modes from their work approach zero as the wavevector $q \to 0$, which is opposite to the scaling relation we derived in this work, and also contradicts the classical theory that low-frequency phonon modes are not likely to scattered. The possible reason is that a too small cutoff of the third-order anharmonic force constants is chosen in their work. If no translational invariance is imposed, the low-frequency phonons are unphysically severely scattered, and thermal conductivity tends to be underestimated.[47]

Finally we note that the observed logarithmic length dependence of the thermal conductivity in silicene might not hold when higher-order phonon scatterings are taken into account since our calculations and analysis only consider the three-phonon process. It is worthwhile to further study the dependence of thermal conductivity on sample size in two-dimensional materials at high temperature, where higher-order phonon scattering might play a role in some of the 2-D materials. Currently there are no widely-accepted conclusions on the contributions of four-phonon processes to the thermal conductivity of solids. On one hand, Raman scattering experiments clearly show the importance of four-phonon processes on the frequency shift and line width of optical phonons,e.g. Balkanski et al.[74] On the other hand, theoretical estimation of three-dimensional bulk materials by Ecsedy and Klemens,[75] shows that the scattering rate of the four-phonon processes is at least two orders of magnitude smaller than the three-phonon processes even at the temperature as high as 1000 K. Whether the four-phonon processes are important in two-dimensional materials is even more unclear. It is thus very desirable to carry out first-principles calculations on the thermal conductivity arising from both cubic and higher-order anharmonicity. However, such a calculation is by no means trivial[74] due to the limitation of computational power. We have thus considered only the three-phonon processes and expect the results very likely to be valid for a wide range of temperature by looking at the contributions of optical phonons to the total thermal conductivity.

## IV. Conclusions

In conclusion, we use the phonon Boltzmann transport equation with the phonon properties calculated by interatomic force constants from the first-principles to predict the thermal conductivity of silicene and graphene. With a detailed analysis on phonon scattering rate, we have shown that silicene has a much smaller thermal conductivity than graphene. Unlike graphene, where most of the heat is conducted by out-of-plane acoustic phonons, the out-of-plane acoustic phonons contribute to only ~10% of the thermal conductivity in silicene. More importantly, in-plane acoustic phonons make the thermal conductivity of silicene unbounded with the increase of the sample size. The differences in phonon transport in silicene and graphene can be attributed to the buckled atomic structure. The buckled structure in silicene breaks the symmetry selection rule that applies to graphene, making the ZA phonon modes strongly scattered so that they contribute very little to heat transport in silicene. The buckled structure also changes the phonon dispersion curve of the ZA branch from $q^{3/2}$ in graphene to linear, so the most important scattering channel in silicene for long wavelength LA phonon modes, LA->ZA+ZA, becomes not as efficient as in graphene to render a finite value for the intrinsic thermal conductivity. In contrast, we proved analytically that the thermal conductivity increases logarithmically with respect to the sample size when both intrinsic phonon-phonon scattering and boundary scattering are considered. This study might shed some light on the fundamental phonon transport mechanisms in other 2-D materials such as transition metal dichalcogenides. For example, strong length-dependence of the thermal conductivity might be expected in $MoS_2$ and $WS_2$ due to a similarly linear dispersion of ZA phonons in these non-one-atom-thick two-dimensional materials.

**Acknowledgments**: This work is supported by the NSF CAREER award (Grant No. 0846561), AFOSR Thermal Sciences Grant (FA9550-11-1-0109), and AFOSR STTR programs (PI: Dr. Sayan Naha). This work utilized the Janus supercomputer, which is supported by the National Science Foundation (award number CNS-0821794), the University of Colorado Boulder, the University of Colorado Denver, and the National Center for Atmospheric Research. The Janus supercomputer is operated by the University of Colorado Boulder. The authors also would like to thank Dr. Yujie Wei from the Institute of Mechanics at the Chinese Academy of Sciences and Professor Mildred Dresselhaus for many valuable discussions.

**Appendix A**

Instead of the definition of dynamical matrix from Eq. (3), the dynamical matrix can also be defined as[45]

$$C_{\tau\tau'}^{\alpha\beta}(\mathbf{q}) = \frac{1}{\sqrt{M_\tau M_{\tau'}}} \sum_{\mathbf{R'}} \phi_{0\tau,\mathbf{R'}\tau'}^{\alpha\beta} e^{i\mathbf{q}\cdot\mathbf{X}_{\mathbf{R'}\tau'}}, \tag{A1}$$

where $\mathbf{X}_{\mathbf{R'}\tau'}$ is the position of atom $\mathbf{R'}\tau'$. The two definitions of dynamical matrix give the identical phonon dispersion relation.[45] It is more convenient to derive the dispersion relation close to the Gamma point using Eq. (A1) as the definition.

The dynamical matrix of a wavevector $\mathbf{q} = q\mathbf{u}$ ($\mathbf{u}$ is a unit vector in the $x$-$y$ plane) can be calculated from the dynamical matrix of its neighbor wavevector $\mathbf{q}_0$ ($\mathbf{q} = \mathbf{q}_0 + \Delta\mathbf{q}$) as a perturbation

$$\begin{aligned}
C_{\tau\tau'}^{\alpha\beta}(\mathbf{q}) &= \frac{1}{\sqrt{M_\tau M_{\tau'}}} \sum_{\mathbf{R'}} \phi_{0\tau,\mathbf{R'}\tau'}^{\alpha\beta} e^{i\mathbf{q}\cdot\mathbf{X}_{\mathbf{R'}\tau'}} = \frac{1}{\sqrt{M_\tau M_{\tau'}}} \sum_{\mathbf{R'}} \phi_{0\tau,\mathbf{R'}\tau'}^{\alpha\beta} e^{i(\mathbf{q}_0+\Delta\mathbf{q})\cdot\mathbf{X}_{\mathbf{R'}\tau'}} \\
&= \frac{1}{\sqrt{M_\tau M_{\tau'}}} \sum_{\mathbf{R'}} \left\{ \phi_{0\tau,\mathbf{R'}\tau'}^{\alpha\beta} e^{i\mathbf{q}_0\cdot\mathbf{X}_{\mathbf{R'}\tau'}} + i(\Delta\mathbf{q}\cdot\mathbf{X}_{\mathbf{R'}\tau'}) \phi_{0\tau,\mathbf{R'}\tau'}^{\alpha\beta} e^{i\mathbf{q}_0\cdot\mathbf{X}_{\mathbf{R'}\tau'}} \right. \\
&\quad \left. -\frac{1}{2}(\Delta\mathbf{q}\cdot\mathbf{X}_{\mathbf{R'}\tau'})^2 \phi_{0\tau,\mathbf{R'}\tau'}^{\alpha\beta} e^{i\mathbf{q}_0\cdot\mathbf{X}_{\mathbf{R'}\tau'}} + ... \right\}
\end{aligned} \tag{A2}$$

If $\mathbf{q}_0$ is chosen as the Gamma point, Eq. (A2) can be simplified as

$$C_{\tau\tau'}^{\alpha\beta}(\mathbf{q}) = \frac{1}{\sqrt{M_\tau M_{\tau'}}} \sum_{\mathbf{R'}} \left\{ \phi_{0\tau,\mathbf{R'}\tau'}^{\alpha\beta} + i(\mathbf{q}\cdot\mathbf{X}_{\mathbf{R'}\tau'}) \phi_{0\tau,\mathbf{R'}\tau'}^{\alpha\beta} - \frac{1}{2}(\mathbf{q}\cdot\mathbf{X}_{\mathbf{R'}\tau'})^2 \phi_{0\tau,\mathbf{R'}\tau'}^{\alpha\beta} + \ldots \right\}$$

$$= C_{\tau\tau'}^{\alpha\beta}(\mathbf{0}) + iq \frac{1}{\sqrt{M_\tau M_{\tau'}}} \sum_{\mathbf{R'}} (\mathbf{u}\cdot\mathbf{X}_{\mathbf{R'}\tau'}) \phi_{0\tau,\mathbf{R'}\tau'}^{\alpha\beta} \quad (A3)$$

$$-q^2 \frac{1}{\sqrt{M_\tau M_{\tau'}}} \sum_{\mathbf{R'}} \left\{ \frac{1}{2}(\mathbf{u}\cdot\mathbf{X}_{\mathbf{R'}\tau'})^2 \phi_{0\tau,\mathbf{R'}\tau'}^{\alpha\beta} \right\} + \ldots$$

$$= C_{\tau\tau'}^{\alpha\beta}(\mathbf{0}) + iq C_{\tau\tau'}^{(1)\alpha\beta} - q^2 C_{\tau\tau'}^{(2)\alpha\beta} + \ldots,$$

where $C^{(1)}$ and $C^{(2)}$ are real matrices. Then the eigenvalues are written as

$$\omega_{\mathbf{q}s}^2 = \omega_{\mathbf{0}s}^2 + iq\left(\varepsilon_{0s}^{\tau\alpha}\right)^* C_{\tau\tau'}^{(1)\alpha\beta} \varepsilon_{0s}^{\tau'\beta} - q^2 \left(\varepsilon_{0s}^{\tau\alpha}\right)^* C_{\tau\tau'}^{(2)\alpha\beta} \varepsilon_{0s}^{\tau'\beta} + \ldots \quad (A4)$$

where $\varepsilon$ is eigenvector of the dynamical matrix $C$, and is real at $\mathbf{q}=0$. Since the eigenvalues of any dynamical matrix are always real, the second term in Eq. (A4) naturally vanishes. Due to the crystal symmetry of silicene, it is easy to prove that $C_{\tau\tau'}^{(2)xz} = C_{\tau\tau'}^{(2)yz} = 0$. Therefore, there is an acoustic branch where the corresponding atomic movement at the Gamma point is pure out-of-plane motion (along $z$ direction), or the ZA branch which is discussed in the main text. The components of its eigenvector are $\varepsilon_{0,ZA}^{\tau z} = \sqrt{2}/2$, $\varepsilon_{0,ZA}^{\tau x} = \varepsilon_{0,ZA}^{\tau y} = 0$ at the Gamma point. The dispersion of this branch becomes quadratic only if the $q^2$ term in Eq. (A4) is zero, or

$$\sum_{\tau'} C_{\tau\tau'}^{(2)zz} = 0, \quad (A5)$$

which requires

$$(\mathbf{u}\cdot\mathbf{X}_{\mathbf{R'}\tau'})^2 \phi_{0\tau,\mathbf{R'}\tau'}^{zz} = 0. \quad (A6)$$

Physically, Eq. (A6) indicates the energy of the 2-D crystal unchanged when every atom move a distance of $\theta\cdot(\mathbf{u}\cdot\mathbf{X}_{\mathbf{R'}\tau'})$ along $z$ direction, where $\theta$ can be regarded as a small angle that the atomic plane rotates, as illustrated in Fig. A1. The deformation due to the shear leads to a relative displacement among atoms. Since the original silicene structure is the structure with the minimum energy, the distorted lattice structure has to be of higher energy. Therefore, Eq. (A6) does not hold for silicene, and there is always a linear component in the ZA dispersion. Such proof provided here can also be applied to other 2-D materials, such as transition metal dichalcogenides.

**Appendix B**

Although the quadratic dispersion can be obtained from first-principles calculations and widely used for the flexural ZA phonons in graphene, some authors have recently pointed out that the quadratic shape of the ZA modes would lead to strong scatterings with long-wavelength LA and TA modes, making the phonon relaxation time of the LA and TA modes approach a constant for phonon near the zone center $q \to 0$.[9, 56] Since the condition for the existence of phonons as elementary excitations is $\omega_{\mathbf{q}s}\tau_{\mathbf{q}s} > 1$,[72] the abnormal constant relaxation times of the LA and TA modes raise a concern on whether the phonon concept can even be applied to the long-wavelength acoustic vibrations in graphene. The problem becomes more severe when the sample size is larger than 1 μm, since the condition $\omega_{\mathbf{q}s}\tau_{\mathbf{q}s} < 1$ occurs in the region of the first Brillouin zone where $q < 5 \times 10^{-5}(2\pi/a_0)$, i.e., where the wavelength $\lambda$ > 1 μm.[9, 56] One possible way to avoid such a concern is to take into account the coupling of ZA and in-plane acoustic phonon modes, in which case the long-wavelength ZA dispersion has the form $\omega \propto q^{3/2}$.[63] In this work, we slightly modify the ZA dispersion by multiplying the original dispersion relation by the factor $\left[1+(q_c/q)^2\right]^{1/4}$, where $q_c$ is a cutoff wave vector, corresponding to the bending rigidity of $\kappa_0 = 1.68$ eV, similar to what Lindsay and Broido did in ref. [10]. The modified ZA branch is then plotted in Fig. 2(a). It is seen that such modification does not change the dispersion of middle- and short-wavelength phonons, but turns the quadratic dispersion to $\omega \propto q^{3/2}$ near the zone center, making $v \propto q^{1/2}$ and $n^0 \propto q^{-3/2}$ for long wavelength ZA phonons. We have thus used the modified dispersion for flexural ZA modes in the calculation of thermal conductivity and phonon scattering rate of graphene in this work.

**Appendix C**

For a three-phonon process in a two-dimensional material, if the mode $\mathbf{q}s$ is fixed and the branch indices $s'$ and $s''$ of the other two modes, $\mathbf{q}'s'$ and $\mathbf{q}''s''$ ($\mathbf{q}'' = \mathbf{q} \pm \mathbf{q}' + \mathbf{G}$), are given, the wavevector $\mathbf{q}'$ has to be on a loop $t(\mathbf{q}')$ in the first Brillouin zone that is determined by the energy conservation condition $\omega_{\mathbf{q}s} \pm \omega_{\mathbf{q}'s'} = \omega_{(\mathbf{q} \pm \mathbf{q}' + \mathbf{G})s''}$. The two-dimensional integral, Eq. (14), can then be converted to a line integral, which is written as[9]

$$\Gamma_{\mathbf{q}s} \propto \int_{t(\mathbf{q}')} \frac{|V_3|^2 (1 + n^0_{\mathbf{q}'s'} + n^0_{\mathbf{q}''s''})}{\left|\partial\left(\omega_{\mathbf{q}s} - \omega_{\mathbf{q}'s'} - \omega_{\mathbf{q}''s''}\right)/\partial \mathbf{q}'\right|} dt \quad \text{(for decay processes)}, \tag{A7}$$

$$\Gamma_{\mathbf{q}s} \propto \int_{t(\mathbf{q}')} \frac{|V_3|^2 (n^0_{\mathbf{q}''s''} - n^0_{\mathbf{q}'s'})}{\left|\partial\left(\omega_{\mathbf{q}s} + \omega_{\mathbf{q}'s'} - \omega_{\mathbf{q}''s''}\right)/\partial \mathbf{q}'\right|} dt \quad \text{(for annihilation processes)}. \tag{A8}$$

To scatter a long-wavelengh phonon mode $\mathbf{q}s$ on a linear dispersion with two other phonon modes $\mathbf{q}'s'$ and $\mathbf{q}''s''$ which are on a dispersion with $\omega \propto q^{3/2}$, it is easy to show that $q' \propto q^{2/3}$ and $q'' \propto q^{2/3}$ to satisfy the energy conservation condition. As a result, the perimeter of the loop $t(\mathbf{q}')$ is on the of order $q^{2/3}$, $(1+n^0_{\mathbf{q}'s'}+n^0_{\mathbf{q}''s''})$ is on the of order $q^{-1}$ and the gradient is on the of order $q^{1/3}$. Now, the only unknown quantity is the three-phonon scattering matrix element $|V_3|^2$. Take the LA phonon in graphene as an example, Eq. (6) can be written as

$$V_3 \propto \frac{\sum_\tau \sum_{\mathbf{R}'\tau'} \sum_{\mathbf{R}''\tau''} \psi^{xzz}_{0\tau,\mathbf{R}'\tau',\mathbf{R}''\tau''} e^{i\mathbf{q}\cdot(\mathbf{r}_\tau - \mathbf{R}'' - \mathbf{r}_{\tau''})} e^{i\mathbf{q}'\cdot(\mathbf{R}' + \mathbf{r}_{\tau'} - \mathbf{R}_\mathbf{R'} - \mathbf{r}_{\tau''})}}{\left(\omega_{\mathbf{q}s}\omega_{\mathbf{q}'s'}\omega_{\mathbf{q}''s''}\right)^{1/2}} . \quad (A9)$$

By expanding the exponential in powers of $q$, we can get the terms of $1, q^{2/3}, q, q^{4/3}, q^{5/3}, q^2, q^{7/3}, \ldots$ In ref. [9], the authors proved that the coefficients of the first six terms would be zero due to the acoustic summation rule and mirror symmetry of the graphene sheet. Since the first nonzero term is $q^{7/3}$, the scattering matrix element $|V_3|^2$ scales with $q^{5/3}$, and finally the scattering rate due to the LA->ZA+ZA process should follow the relation as $\Gamma_{\mathbf{q}s} \propto q$ as $q \to 0$.

## Appendix D

In Sec. III(D), we discussed three possible scattering mechanisms for a long-wavelength acoustic phonon mode $\mathbf{q}s$ that satisfies the same linear phonon dispersion. One is the decay process that scatters into two long-wavelength acoustic modes $\mathbf{q}'s'$ and $\mathbf{q}''s''$. The other two are the annihilation scattering with a high-frequency mode $\mathbf{q}'s'$ and generate another high-frequency mode $\mathbf{q}''s''$ that could be on the same branch as $\mathbf{q}'s'$ or on a different branch.

In the first case, all of the three long-wavelength modes are on the linear dispersion curves, and the scattering matrix can be derived from a continuum model, which gives $|V_3|^2 \propto qq'q''$.[51] Since both $q'$ and $q''$ are in the order of $q$ (otherwise energy conservation condition cannot be satisfied), we can get these scaling relations: $|V_3|^2 \propto q^3$, $\left(1+n^0_{\mathbf{q}'s'}+n^0_{\mathbf{q}''s''}\right) \propto q^{-1}$ and $\left|\nabla_{\mathbf{q}'}\left(\omega_{\mathbf{q}s}-\omega_{\mathbf{q}'s'}-\omega_{\mathbf{q}''s''}\right)\right| \propto 1$ through some algebra. In addition, the perimeter of the loop $t(\mathbf{q}')$ is in the same order of $q'$. Therefore, the scattering rate $\Gamma_{\mathbf{q}s}$ due to the decay processes into two long-wavelength modes is in the order of $q^3$ according to Eq. (A7).

In the other two cases, $\mathbf{q}'s'$ and $\mathbf{q}''s''$ are not long-wavelength modes. The scattering matrix element $|V_3|^2$ becomes proportional to $q$ and $\left(n^0_{\mathbf{q}''s''}-n^0_{\mathbf{q}'s'}\right)$ can be estimated by $n^0\left(\omega_{\mathbf{q}''s''}\right)-n^0\left(\omega_{\mathbf{q}'s'}\right)=\left[\partial n^0\left(\omega_{\mathbf{q}'s'}\right)/\partial \omega_{\mathbf{q}'s'}\right]vq \propto q$, which $v_{\mathbf{q}s}q$ is the energy of $\mathbf{q}s$. If modes $\mathbf{q}'s'$ and $\mathbf{q}''s''$ are on the same branch, the gradient $\left|\nabla_{\mathbf{q}'}\left(\omega_{\mathbf{q}s}-\omega_{\mathbf{q}'s'}+\omega_{\mathbf{q}''s''}\right)\right|$ can be estimated by $\left|\left(\partial \omega_{\mathbf{q}'s'}/\partial \mathbf{q}'\right)\cdot \mathbf{q}\right| \propto q$ and the perimeter of the loop $t(\mathbf{q}')$ is a finite value. Then according to Eq. (A8) we can obtain $\Gamma_{\mathbf{q}s} \propto q$ for the second case. If $\mathbf{q}'s'$ and $\mathbf{q}''s''$ are not on the same branch, $\left|\nabla_{\mathbf{q}'}\left(\omega_{\mathbf{q}s}-\omega_{\mathbf{q}'s'}+\omega_{\mathbf{q}''s''}\right)\right|=\left|\mathbf{v}_{\mathbf{q}'s'}-\mathbf{v}_{\mathbf{q}''s''}\right|$ is a finite number. Now, $\Gamma_{\mathbf{q}s} \propto q^2$ for the third case.

## References

1   A. C. Neto, F. Guinea, N. Peres, K. S. Novoselov, and A. K. Geim, Rev Mod Phys **81**, 109 (2009).
2   A. K. Geim, science **324**, 1530 (2009).
3   A. K. Geim and K. S. Novoselov, Nature Mater **6**, 183 (2007).
4   A. A. Balandin, Nature Mater **10**, 569 (2011).

5   E. Pop, V. Varshney, and A. K. Roy, Mrs Bull **37**, 1273 (2012).
6   D. L. Nika and A. A. Balandin, Journal of Physics: Condensed Matter **24**, 233203 (2012).
7   D. Singh, J. Y. Murthy, and T. S. Fisher, J Appl Phys **110**, 044317 (2011).
8   L. F. C. Pereira and D. Donadio, Phys Rev B **87**, 125424 (2013).
9   N. Bonini, J. Garg, and N. Marzari, Nano Lett **12**, 2673 (2012).
10  L. Lindsay, D. Broido, and N. Mingo, Phys Rev B **82**, 115427 (2010).
11  X. Li, K. Maute, M. L. Dunn, and R. Yang, Phys Rev B **81**, 245318 (2010).
12  D. Nika, E. Pokatilov, A. Askerov, and A. Balandin, Physical Review B **79**, 155413 (2009).
13  K. M. Shahil and A. A. Balandin, Nano letters **12**, 861 (2012).
14  V. Goyal and A. A. Balandin, Applied Physics Letters **100**, 073113 (2012).
15  P. Goli, S. Legedza, A. Dhar, R. Salgado, J. Renteria, and A. A. Balandin, Journal of Power Sources **248**, 37 (2014).
16  Z. Yan, G. Liu, J. M. Khan, and A. A. Balandin, Nature communications **3**, 827 (2012).
17  H. Malekpour, K.-H. Chang, J.-C. Chen, C.-Y. Lu, D. Nika, K. S. Novoselov, and A. A. Balandin, Nano letters **14**, 5155 (2014).
18  S. Z. Butler, et al., ACS nano **7**, 2898 (2013).
19  M. Xu, T. Liang, M. Shi, and H. Chen, Chemical reviews **113**, 3766 (2013).
20  S. Cahangirov, M. Topsakal, E. Aktürk, H. Şahin, and S. Ciraci, Phys Rev Lett **102**, 236804 (2009).
21  A. Fleurence, R. Friedlein, T. Ozaki, H. Kawai, Y. Wang, and Y. Yamada-Takamura, Phys Rev Lett **108**, 245501 (2012).
22  P. Vogt, P. De Padova, C. Quaresima, J. Avila, E. Frantzeskakis, M. C. Asensio, A. Resta, B. Ealet, and G. Le Lay, Phys Rev Lett **108**, 155501 (2012).
23  B. Wang, J. Wu, X. Gu, H. Yin, Y. Wei, R. Yang, and M. Dresselhaus, Appl Phys Lett **104**, 081902 (2014).
24  L. Chen, C.-C. Liu, B. Feng, X. He, P. Cheng, Z. Ding, S. Meng, Y. Yao, and K. Wu, Phys Rev Lett **109**, 056804 (2012).
25  N. Drummond, V. Zolyomi, and V. Fal'Ko, Phys Rev B **85**, 075423 (2012).
26  Z. Ni, Q. Liu, K. Tang, J. Zheng, J. Zhou, R. Qin, Z. Gao, D. Yu, and J. Lu, Nano letters **12**, 113 (2011).

27 X. Zhang, H. Xie, M. Hu, H. Bao, S. Yue, G. Qin, and G. Su, Phys Rev B **89**, 054310 (2014).

28 M. Hu, X. Zhang, and D. Poulikakos, Phys Rev B **87**, 195417 (2013).

29 X. Xu, et al., Nature Communications **5**, 3689 (2014).

30 Q.-X. Pei, Y.-W. Zhang, Z.-D. Sha, and V. B. Shenoy, J Appl Phys **114**, 033526 (2013).

31 T. Y. Ng, J. Yeo, and Z. Liu, International Journal of Mechanics and Materials in Design **9**, 105 (2013).

32 A. Ward, D. Broido, D. A. Stewart, and G. Deinzer, Phys Rev B **80**, 125203 (2009).

33 K. Esfarjani, G. Chen, and H. T. Stokes, Physical Review B **84**, 085204 (2011).

34 T. Luo, J. Garg, J. Shiomi, K. Esfarjani, and G. Chen, EPL (Europhysics Letters) **101**, 16001 (2013).

35 J. Garg, N. Bonini, B. Kozinsky, and N. Marzari, Physical Review Letters **106**, 045901 (2011).

36 Z. Tian, J. Garg, K. Esfarjani, T. Shiga, J. Shiomi, and G. Chen, Physical Review B **85**, 184303 (2012).

37 J. Shiomi, K. Esfarjani, and G. Chen, Phys Rev B **84**, 104302 (2011).

38 W. Li, L. Lindsay, D. Broido, D. A. Stewart, and N. Mingo, Physical Review B **86**, 174307 (2012).

39 J. Garg, N. Bonini, and N. Marzari, Nano letters **11**, 5135 (2011).

40 J. Garg and G. Chen, Physical Review B **87**, 140302 (2013).

41 W. Li, N. Mingo, L. Lindsay, D. Broido, D. Stewart, and N. Katcho, Phys Rev B **85**, 195436 (2012).

42 W. Li and N. Mingo, Journal of Applied Physics **114**, 054307 (2013).

43 L. Paulatto, F. Mauri, and M. Lazzeri, Phys Rev B **87**, 214303 (2013).

44 X. Tang and B. Fultz, Phys Rev B **84**, 054303 (2011).

45 G. P. Srivastava, *The physics of phonons* (CRC Press, 1990).

46 P. Giannozzi, et al., Journal of Physics: Condensed Matter **21**, 395502 (2009).

47 L. Lindsay, D. Broido, and T. Reinecke, Physical Review B **87**, 165201 (2013).

48 A. J. McGaughey and A. Jain, Appl Phys Lett **100**, 061911 (2012).

49 R. Yang and G. Chen, Physical Review B **69**, 195316 (2004).

50 N. Mingo and D. Broido, Nano letters **5**, 1221 (2005).

51  J. M. Ziman, *Electrons and phonons: the theory of transport phenomena in solids* (Oxford University Press, 2001).

52  J. R. Yates, X. Wang, D. Vanderbilt, and I. Souza, Phys Rev B **75**, 195121 (2007).

53  G. Fugallo, M. Lazzeri, L. Paulatto, and F. Mauri, Physical Review B **88**, 045430 (2013).

54  H. A. Van der Vorst, SIAM Journal on scientific and Statistical Computing **13**, 631 (1992).

55  M. Omini and A. Sparavigna, Phys Rev B **53**, 9064 (1996).

56  A. Chernatynskiy and S. R. Phillpot, Phys Rev B **82**, 134301 (2010).

57  E. Scalise, M. Houssa, G. Pourtois, B. van den Broek, V. Afanas'ev, and A. Stesmans, Nano Research **6**, 19 (2013).

58  A. Ferrari, et al., Phys Rev Lett **97**, 187401 (2006).

59  H. Şahin, S. Cahangirov, M. Topsakal, E. Bekaroglu, E. Akturk, R. T. Senger, and S. Ciraci, Phys Rev B **80**, 155453 (2009).

60  X. Li, J. T. Mullen, Z. Jin, K. M. Borysenko, M. B. Nardelli, and K. W. Kim, Phys Rev B **87**, 115418 (2013).

61  H. Zabel, Journal of Physics: Condensed Matter **13**, 7679 (2001).

62  X. Li, J. T. Mullen, Z. Jin, K. M. Borysenko, M. B. Nardelli, and K. W. Kim, Physical Review B **87**, 115418 (2013).

63  E. Mariani and F. von Oppen, Phys Rev Lett **100**, 076801 (2008).

64  L. Lindsay and D. Broido, Phys Rev B **81**, 205441 (2010).

65  S. Chen, Q. Wu, C. Mishra, J. Kang, H. Zhang, K. Cho, W. Cai, A. A. Balandin, and R. S. Ruoff, Nature Mater **11**, 203 (2012).

66  S. Chen, et al., ACS nano **5**, 321 (2010).

67  J.-U. Lee, D. Yoon, H. Kim, S. W. Lee, and H. Cheong, Phys Rev B **83**, 081419 (2011).

68  C. Faugeras, B. Faugeras, M. Orlita, M. Potemski, R. R. Nair, and A. Geim, ACS nano **4**, 1889 (2010).

69  D. Li, Y. Wu, P. Kim, L. Shi, P. Yang, and A. Majumdar, Appl Phys Lett **83**, 2934 (2003).

70  W. Liu and M. Asheghi, Journal of heat transfer **128**, 75 (2006).

71  F. H. Stillinger and T. A. Weber, Phys Rev B **31**, 5262 (1985).

72  D. G. Cahill, et al., Applied Physics Reviews **1**, 011305 (2014).

73  H. Xie, M. Hu, and H. Bao, Applied Physics Letters **104**, 131906 (2014).

74  M. Balkanski, R. Wallis, and E. Haro, Phys Rev B **28**, 1928 (1983).

75  D. Ecsedy and P. Klemens, Phys Rev B **15**, 5957 (1977).

**FIGURE CAPTIONS**

Fig. 1. (color online) (a) Top and side views of the buckled atomic structure of silicene. The parallelogram denotes the primitive unit cell of silicene. (b) First Brillouin zone of silicene (as well as graphene). The black dots are the $q$-grid points used to calculate the thermal conductivity. (c) The schematics of the simulation domain for a finite size L sample.

Fig. 2. (color online) Phonon dispersion of graphene (a) and silicene (b) calculated from the first-principles. The red line in (a) is the renormalized dispersion curve for ZA modes which satisfies $\omega \propto q^{3/2}$ near the zone center. This renormalized dispersion relation is used to calculate and analyze the phonon scattering process and thermal conductivity of graphene in this work.

Fig. 3 (color online) The dependence of the thermal conductivity on $q$-grid for graphene (a) and silicene (b) with sample size $L = 3\mu\text{m}$.

Fig. 4. (color online) Thermal conductivity of graphene as a function of temperature. The solid, dashed, dashed-dotted black lines are the calculated thermal conductivity of graphene with sample sizes $L$=100 μm, 10 μm and 3 μm, respectively. The pink line is the PBTE solution from Singh *et al*'s work [7] for an infinite size sample. Experimental values [65-68] are represented by open symbols: circles[65] represent measured values for isotopically enriched graphene with 99.99% $^{12}$C and 0.01% $^{13}$C, while triangles,[66] diamonds[67] and pentagons[68] are the values for naturally occurring graphene.

Fig. 5. (color online) Calculated thermal conductivity of silicene as a function of temperature.

Fig. 6. (color online) Scaled thermal conductivities from different phonon branches of (a) graphene and (b) silicene with sample size $L=10\mu m$ as a function of temperature.

Fig. 7. (color online) Scattering rates of acoustic phonon modes of graphene (a) and silicene (b) along Γ-M direction at 300 K.

Fig. 8. (color online) Scattering rates of out-of-plane acoustic phonon modes of graphene (a) and silicene (b) along Γ-M direction at 300 K.

Fig. 9. (color online) Length dependence of thermal conductivity of each phonon branch of (a) graphene and (b) silicene. The inset in (b) is the q-dependence of thermal conductivity of silicene when the boundary scattering is not considered.

Fig. 10. (color online) Accumulated thermal conductivity of the LA and TA branches of silicene as a function of phonon frequency. The values are calculated using a 128×128 $q$-grid.

Fig. 11. (color online) Scattering rates of in-plane LA and TA acoustic phonon modes in graphene (a) and silicene (b) along the Γ-M direction at 300 K using log-log scale.

Fig. A1. (color online) The schematic of the shear deformation of silicene.

**Figures**

Figure 1

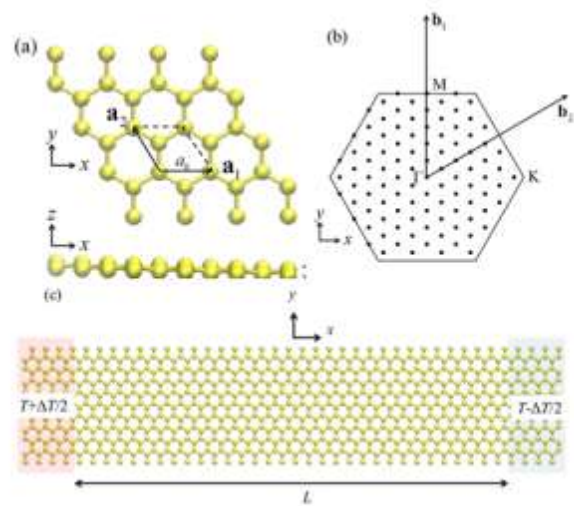

Figure 2

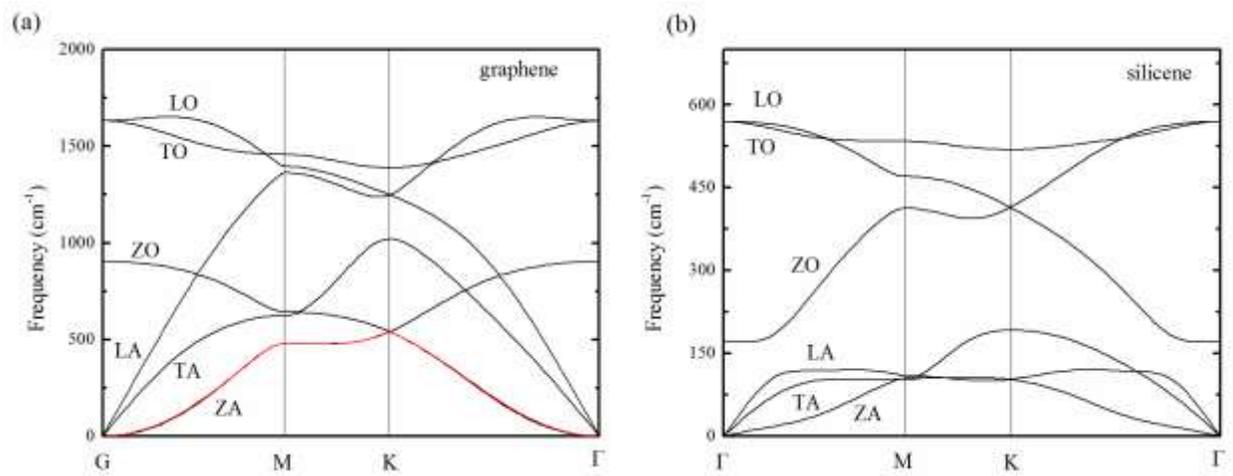

Figure 3

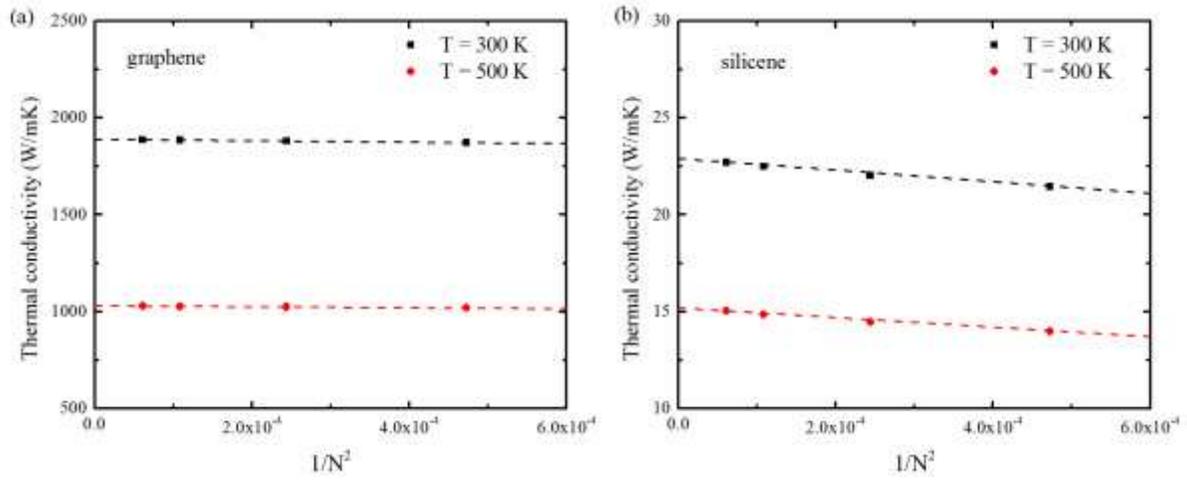

Figure 4

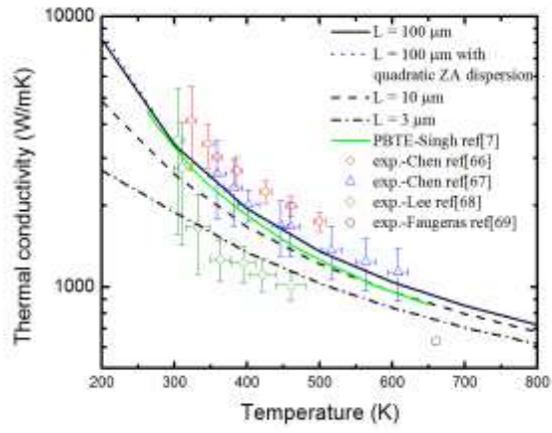

Figure 5

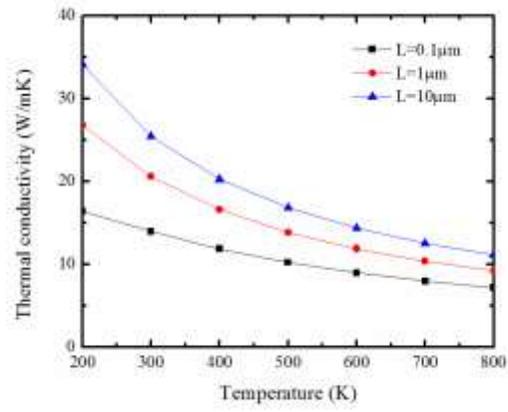

Figure 6

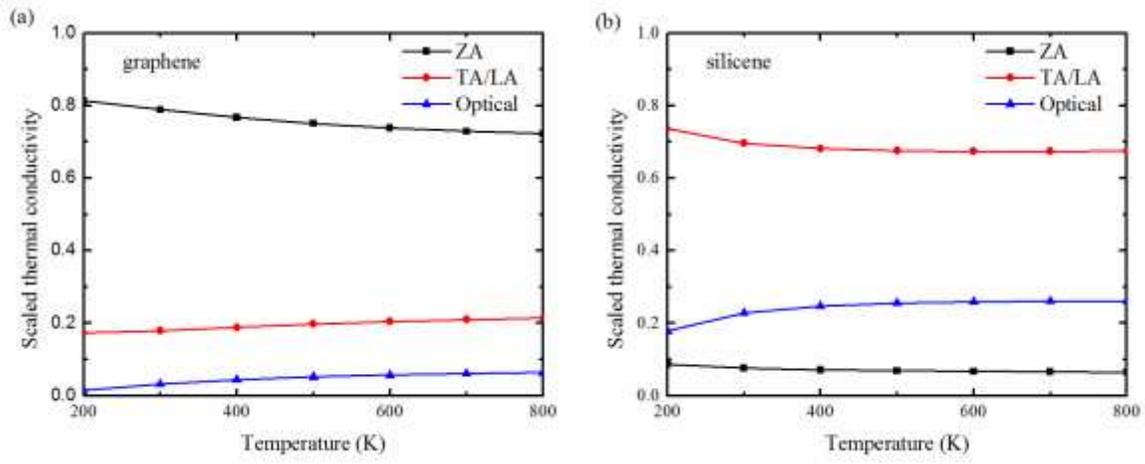

Figure 7

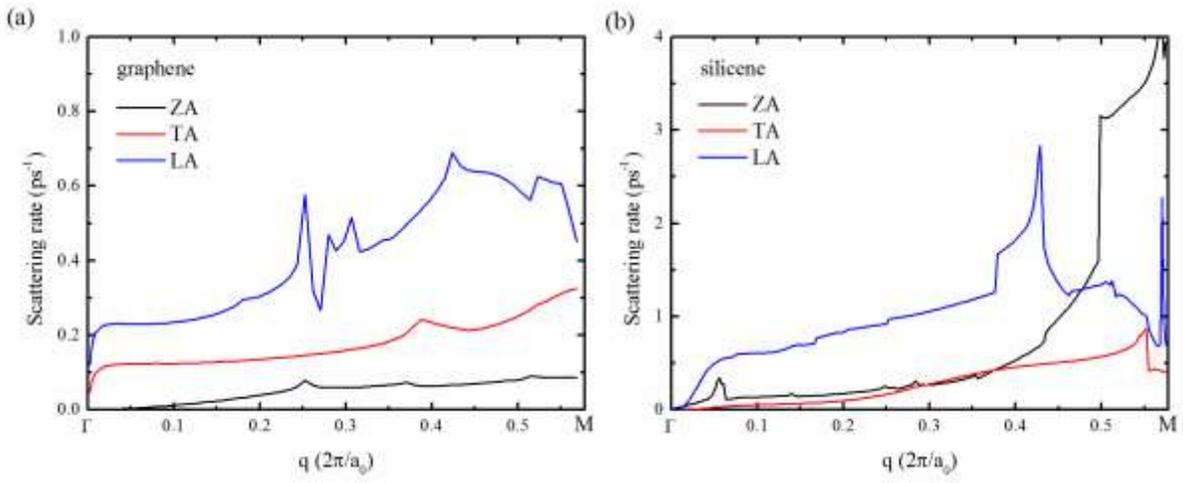

Figure 8

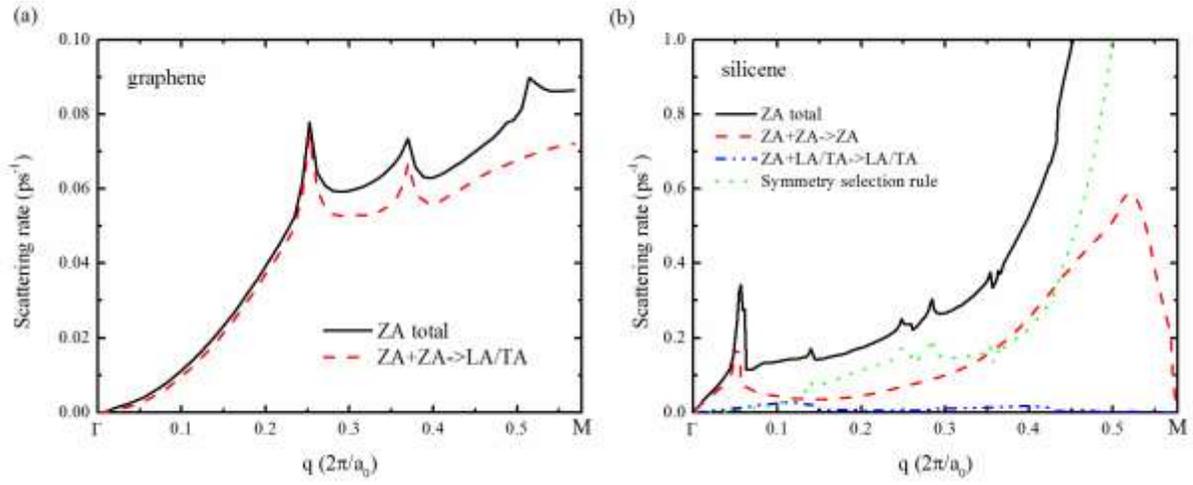

Figure 9

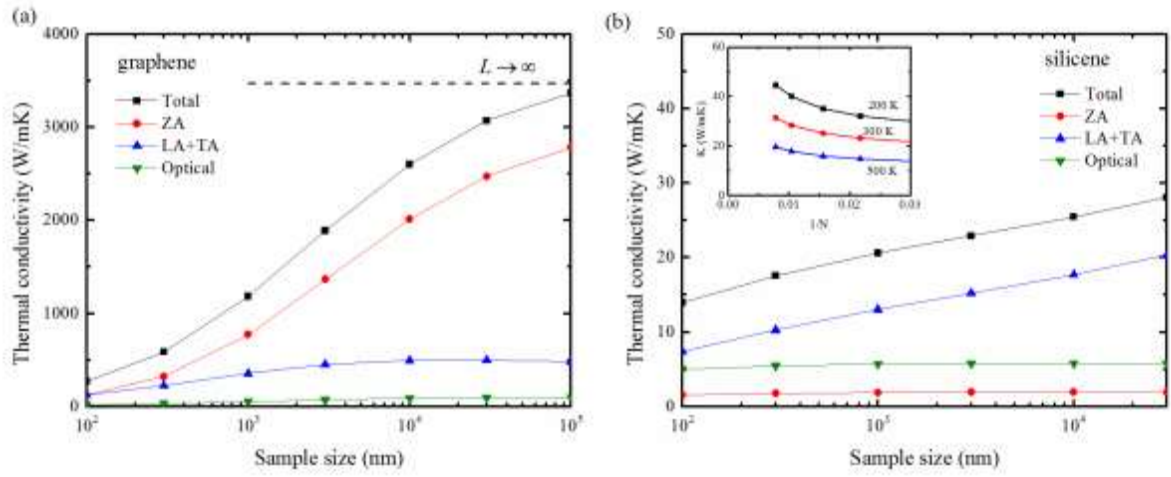

Figure 10

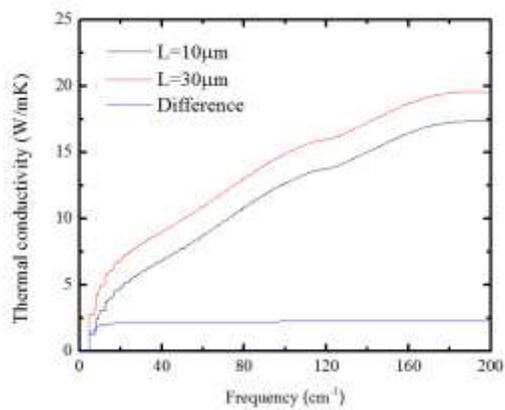

Figure 11

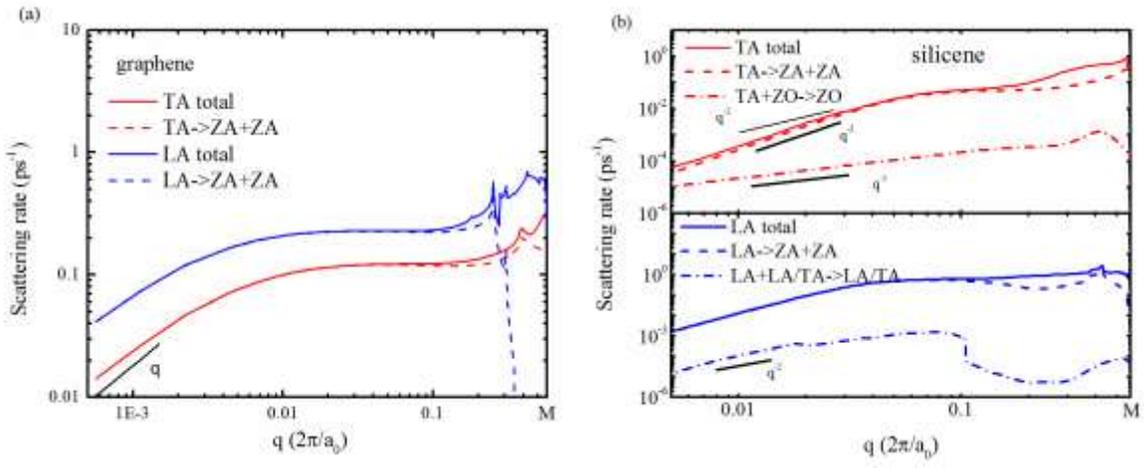

Figure A1

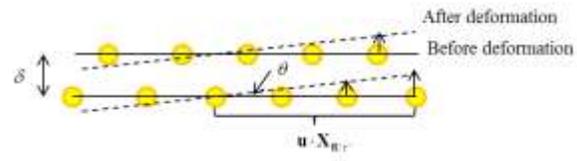

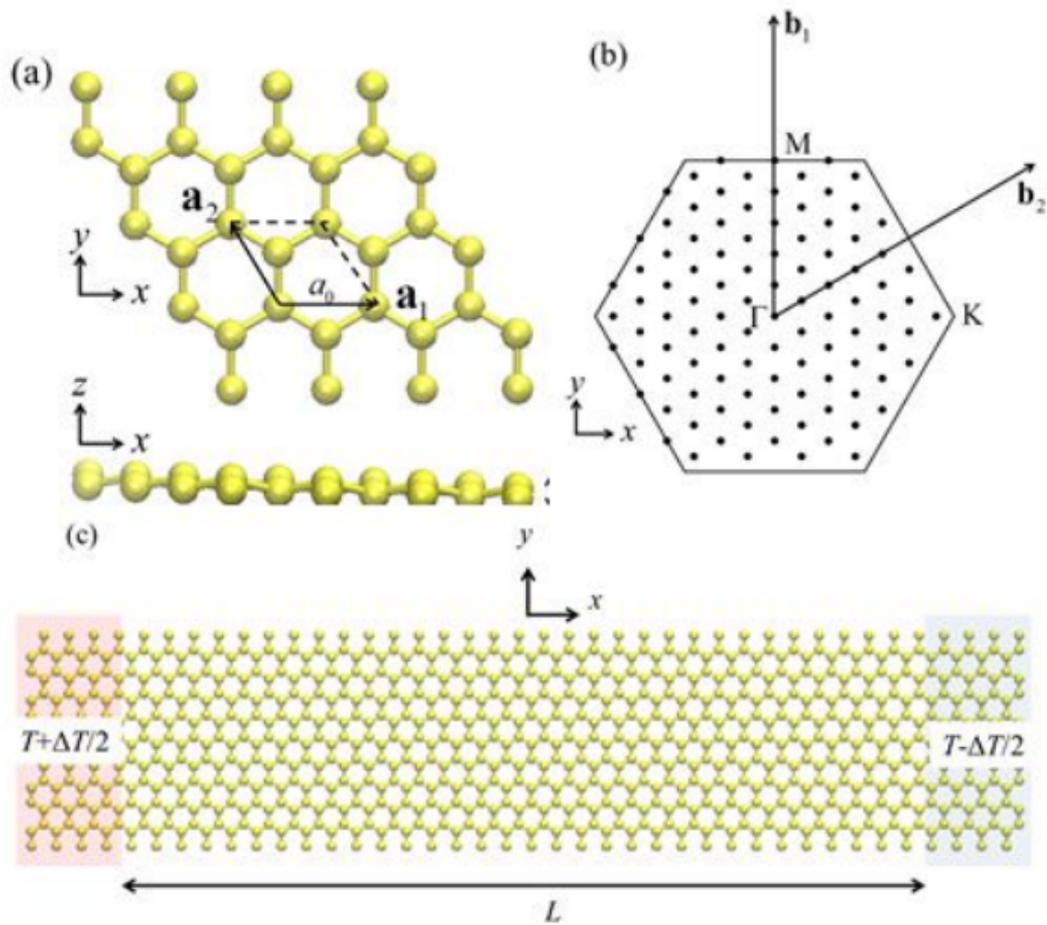

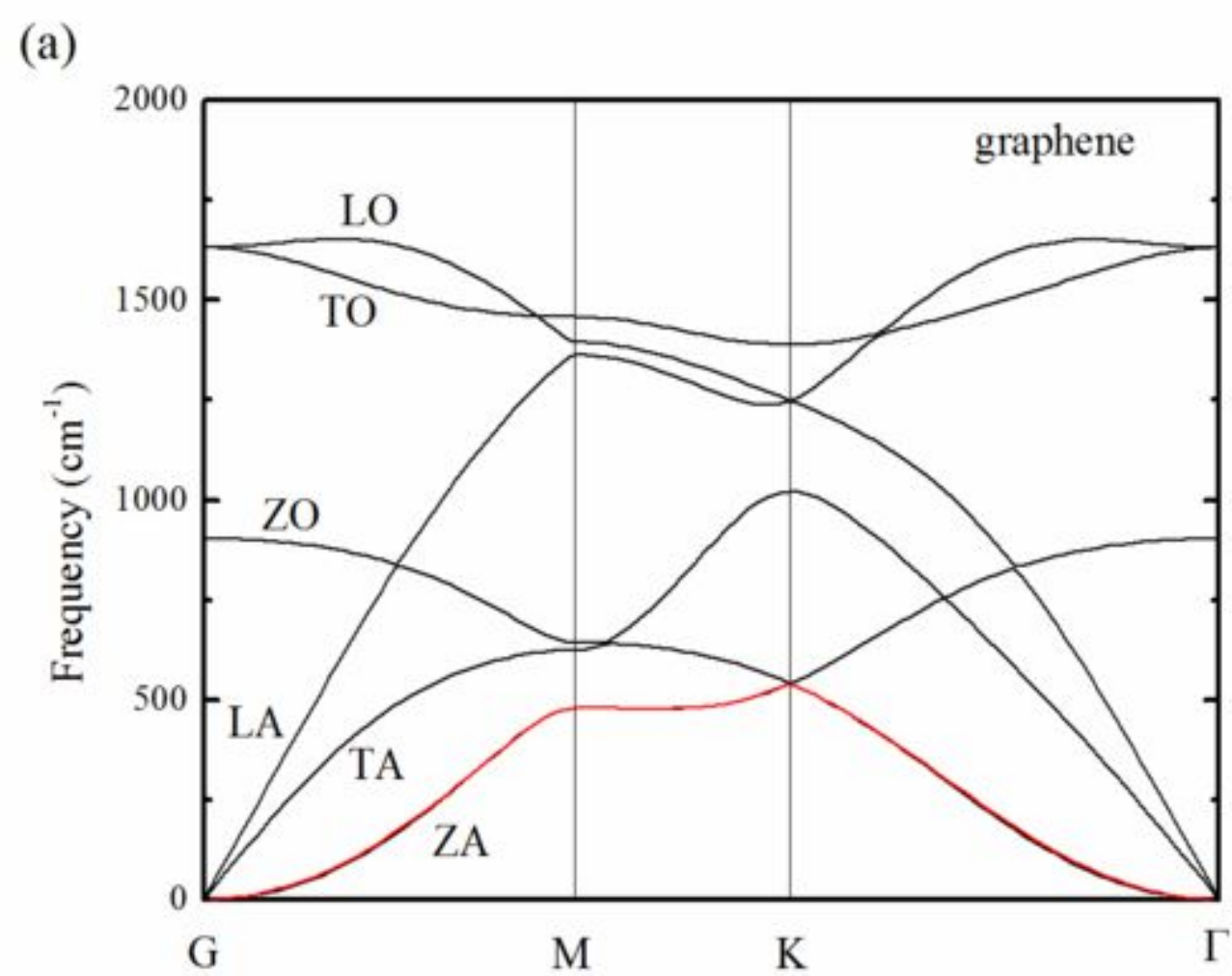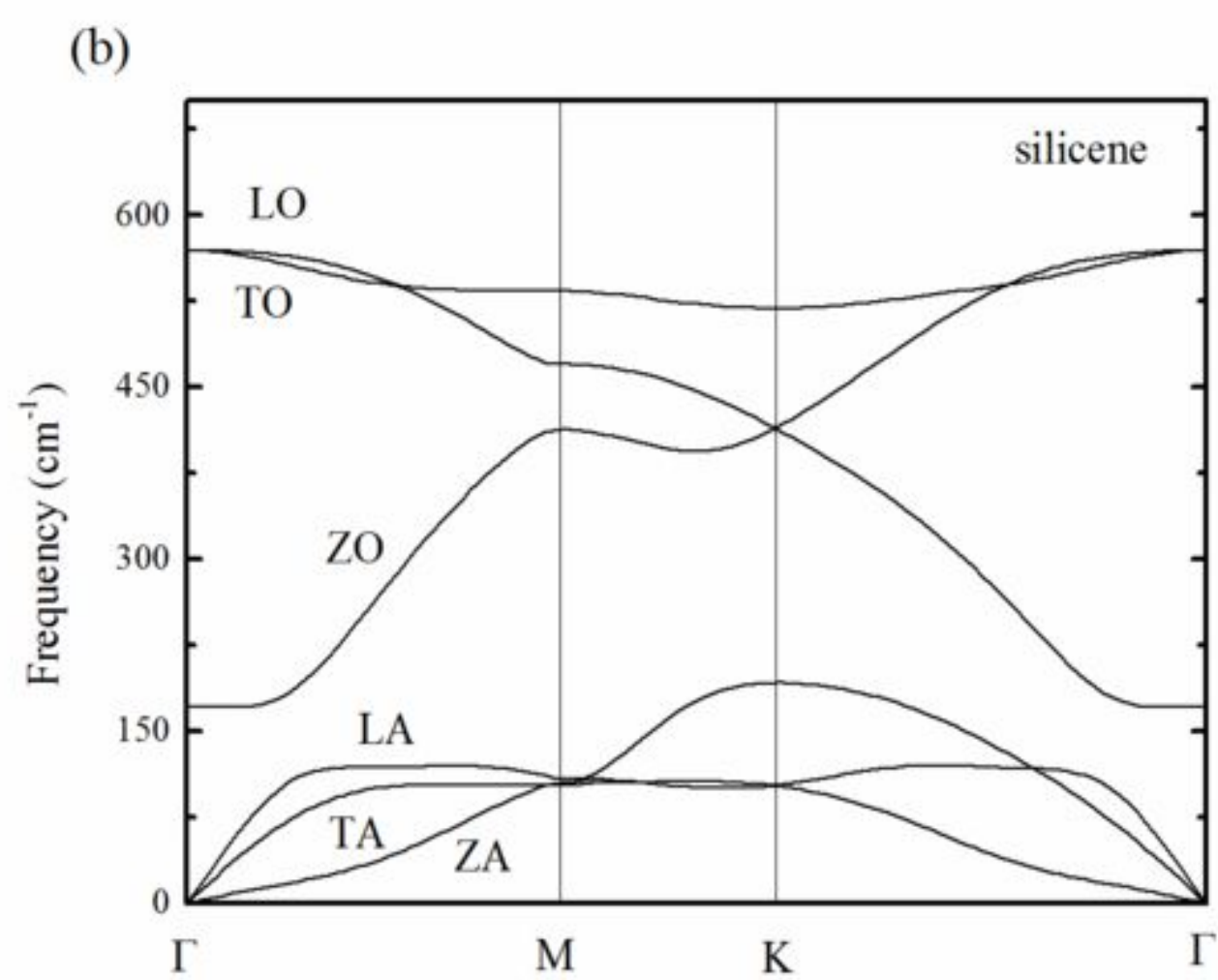

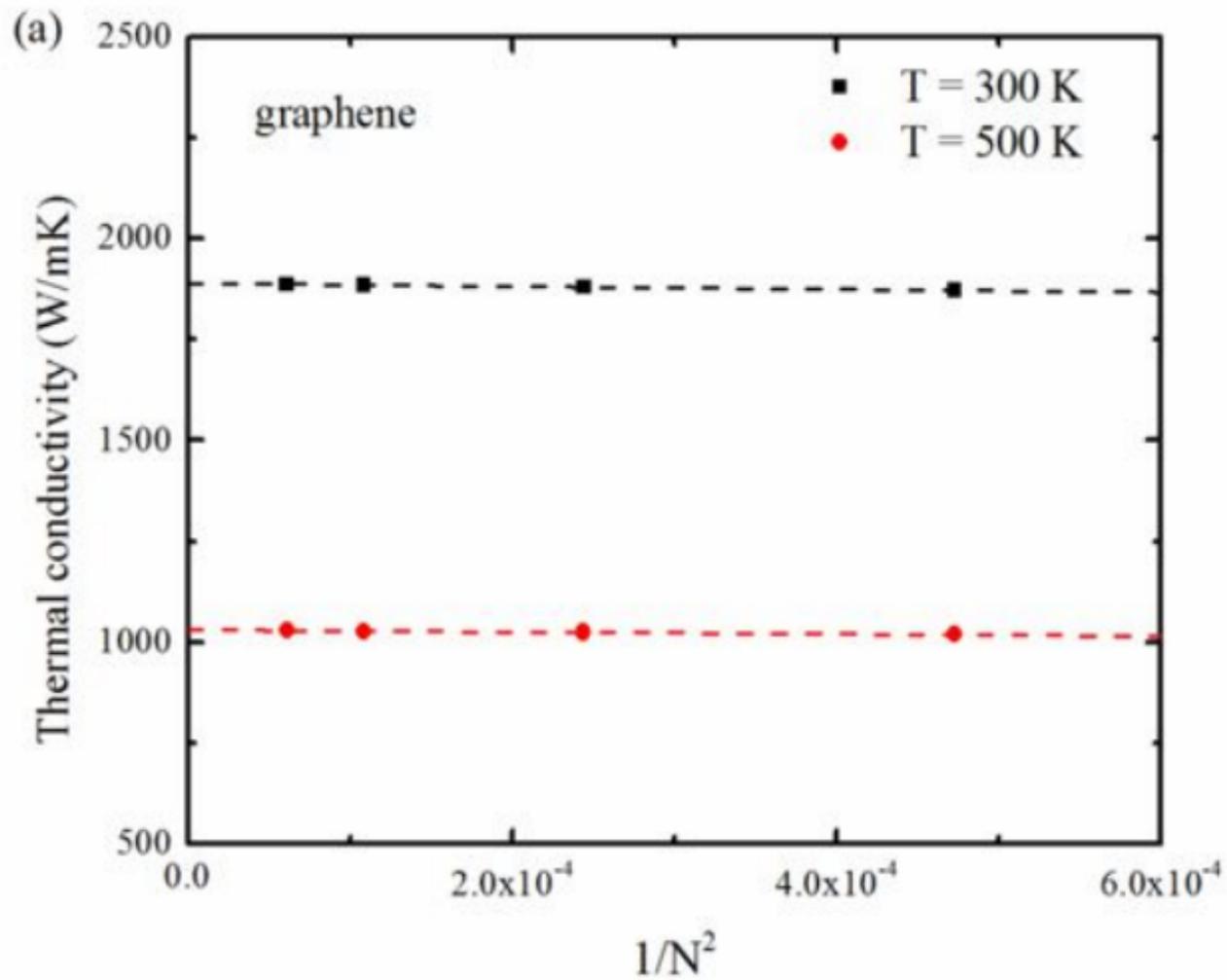 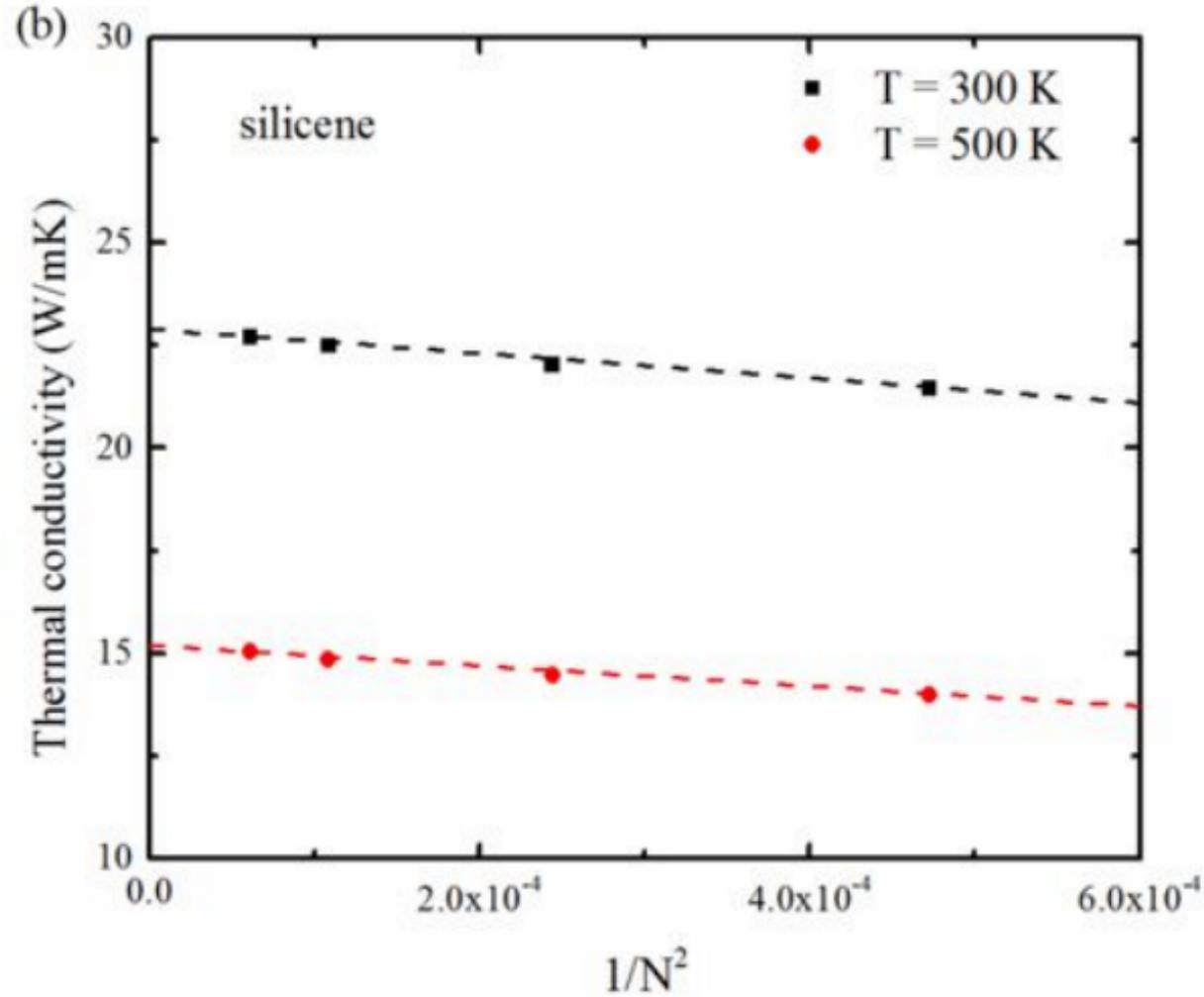

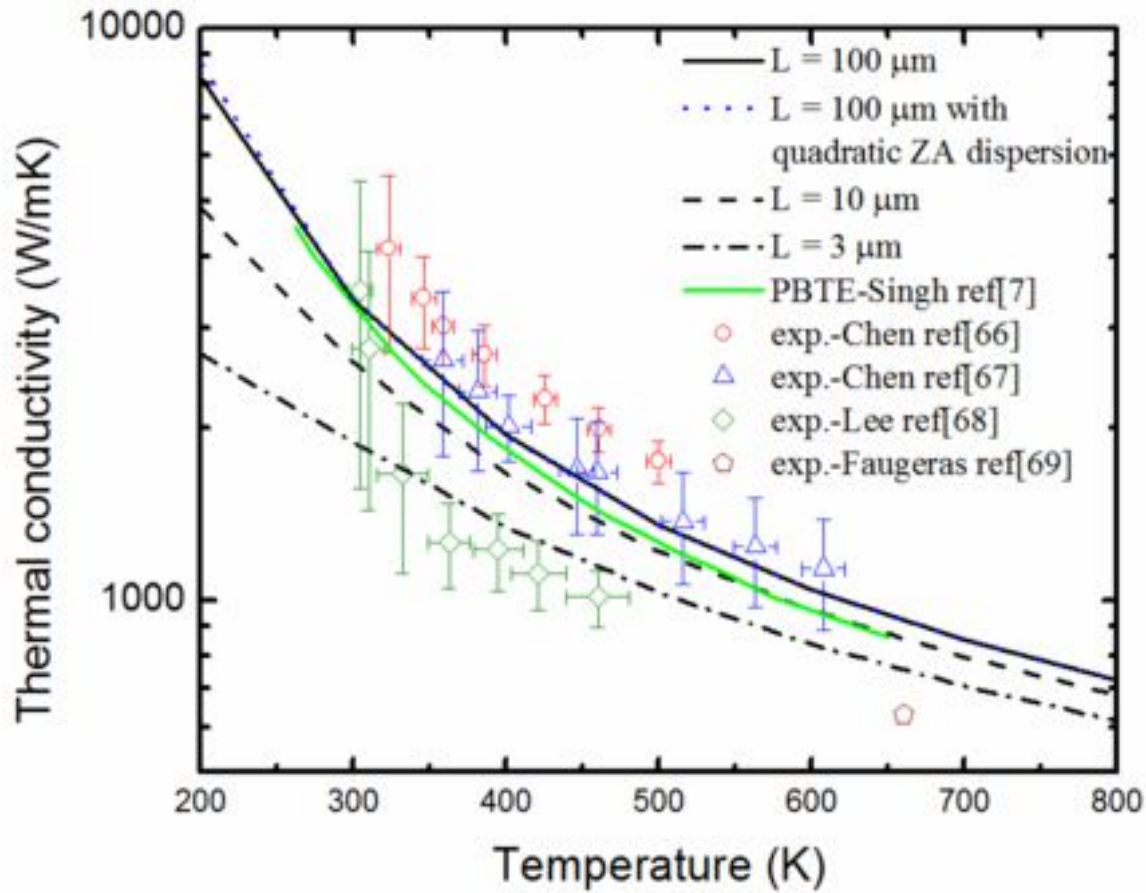

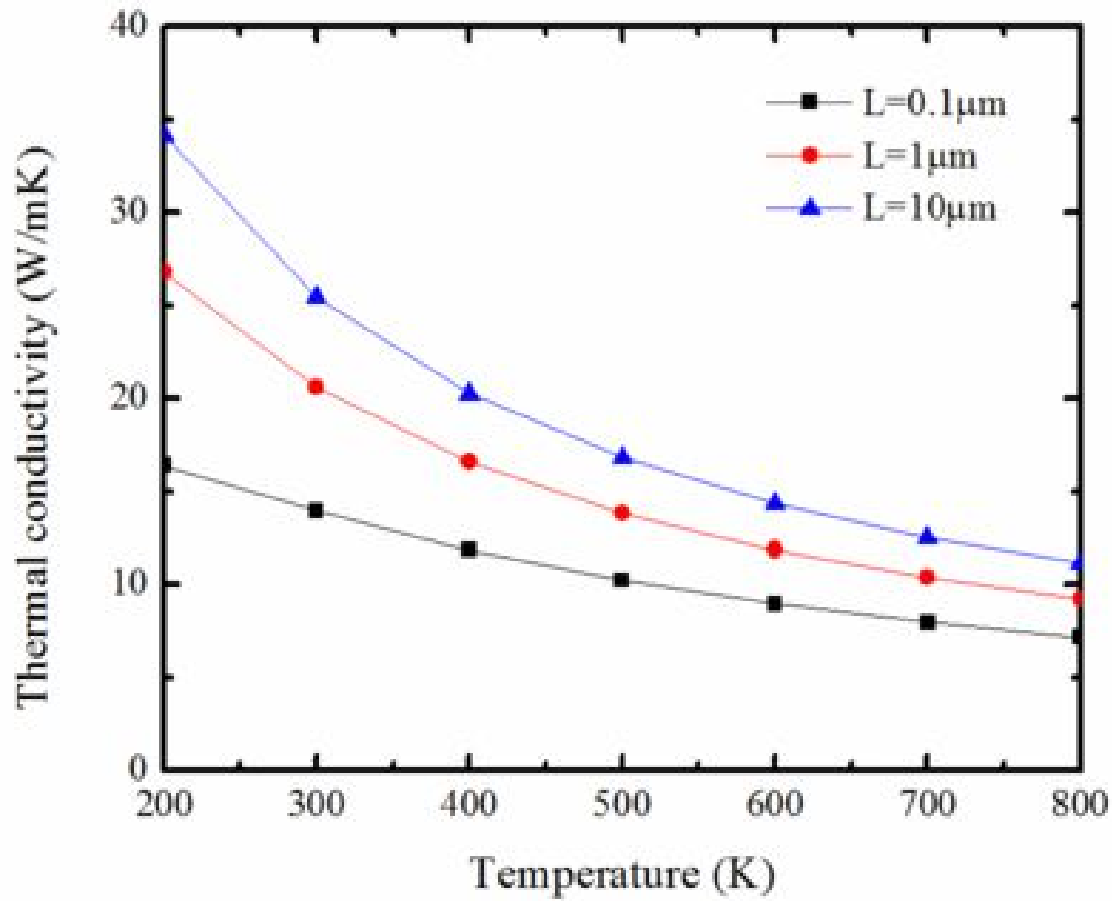

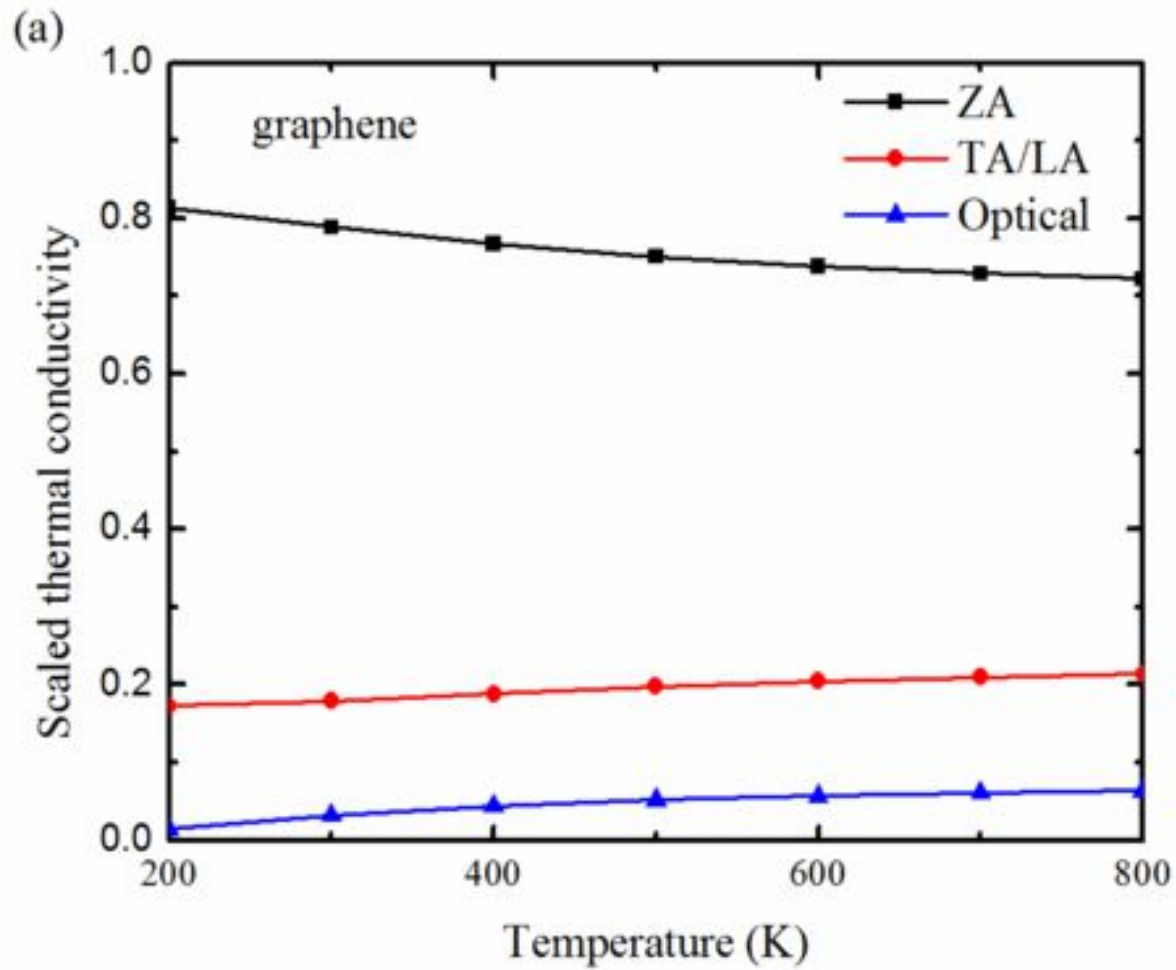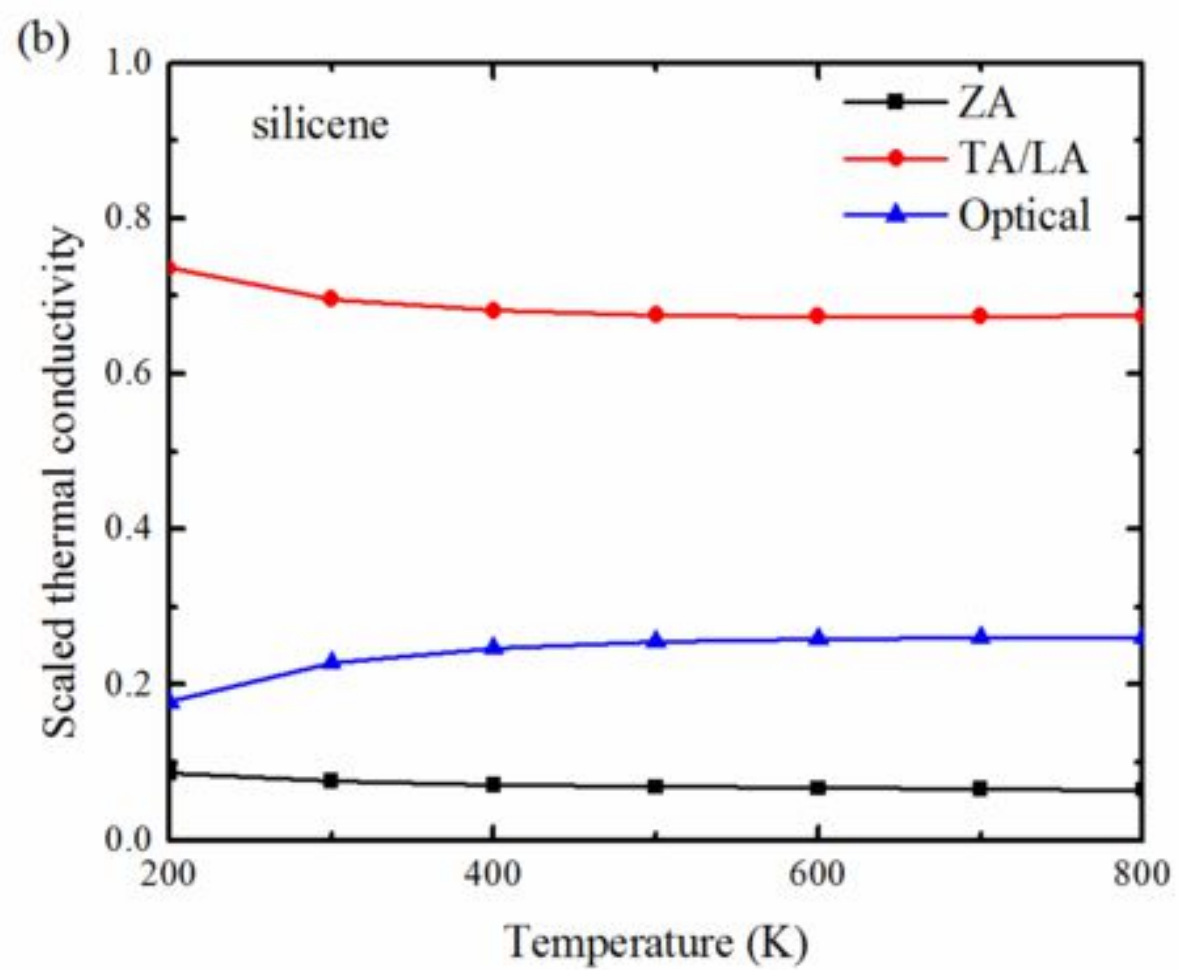

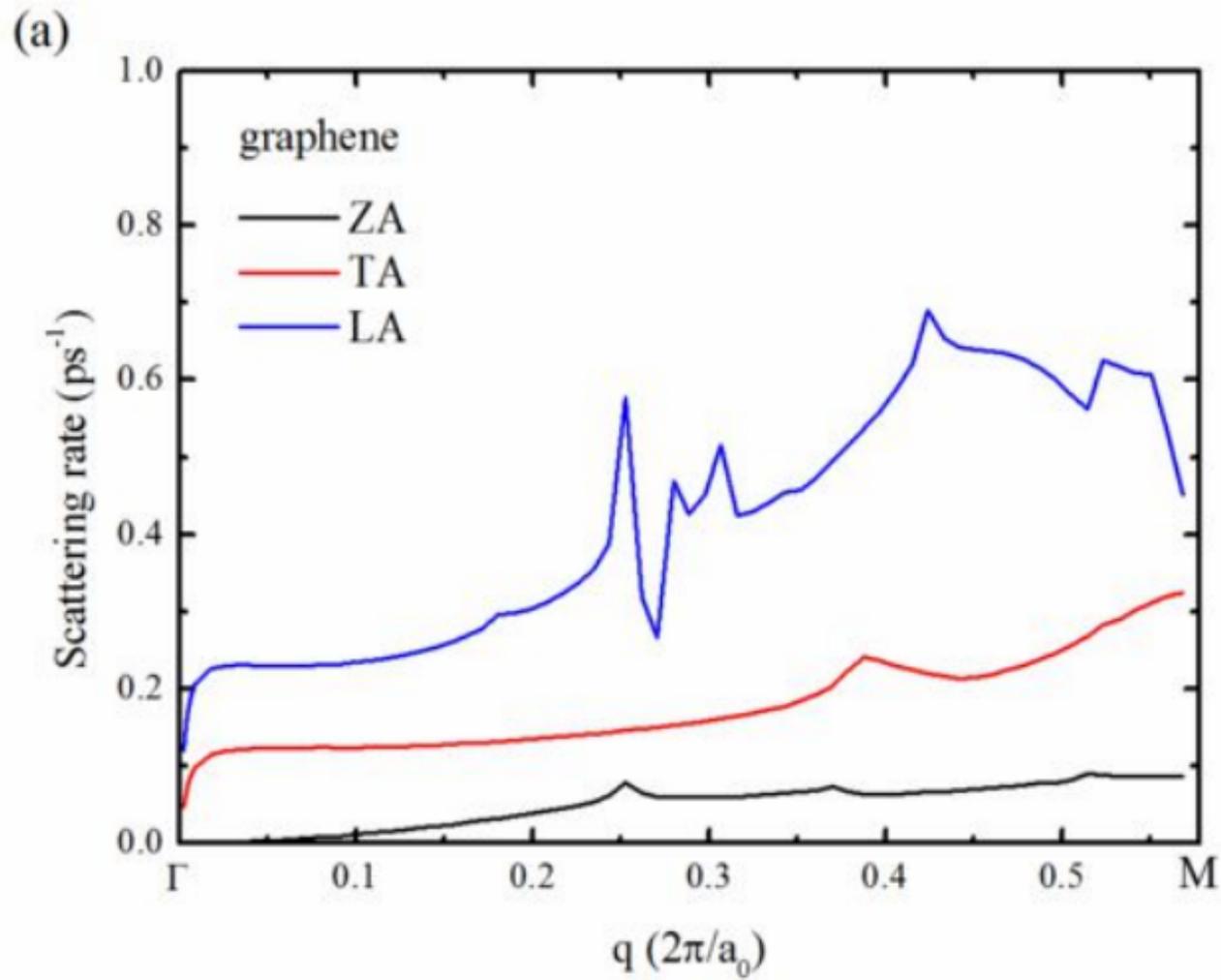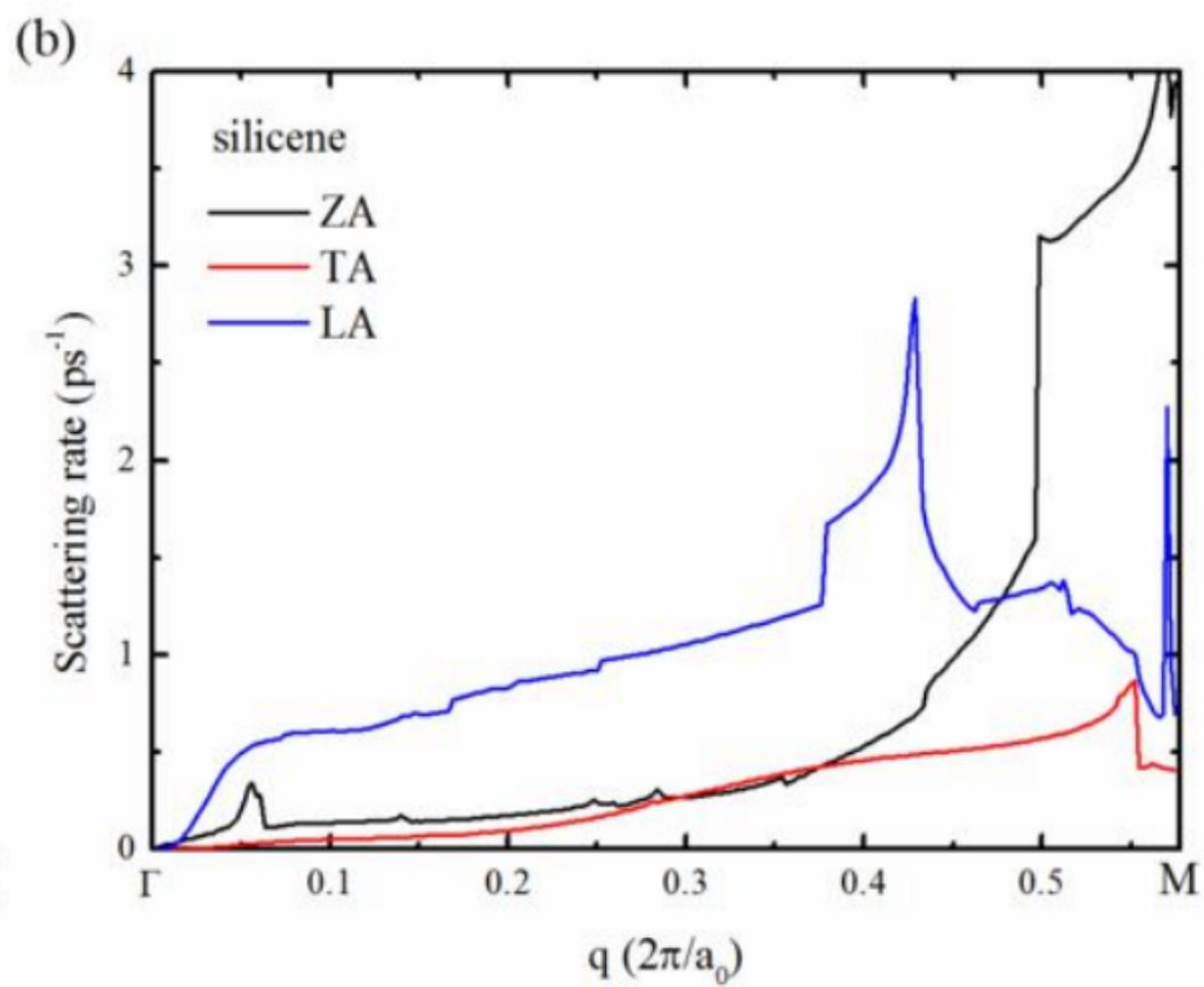

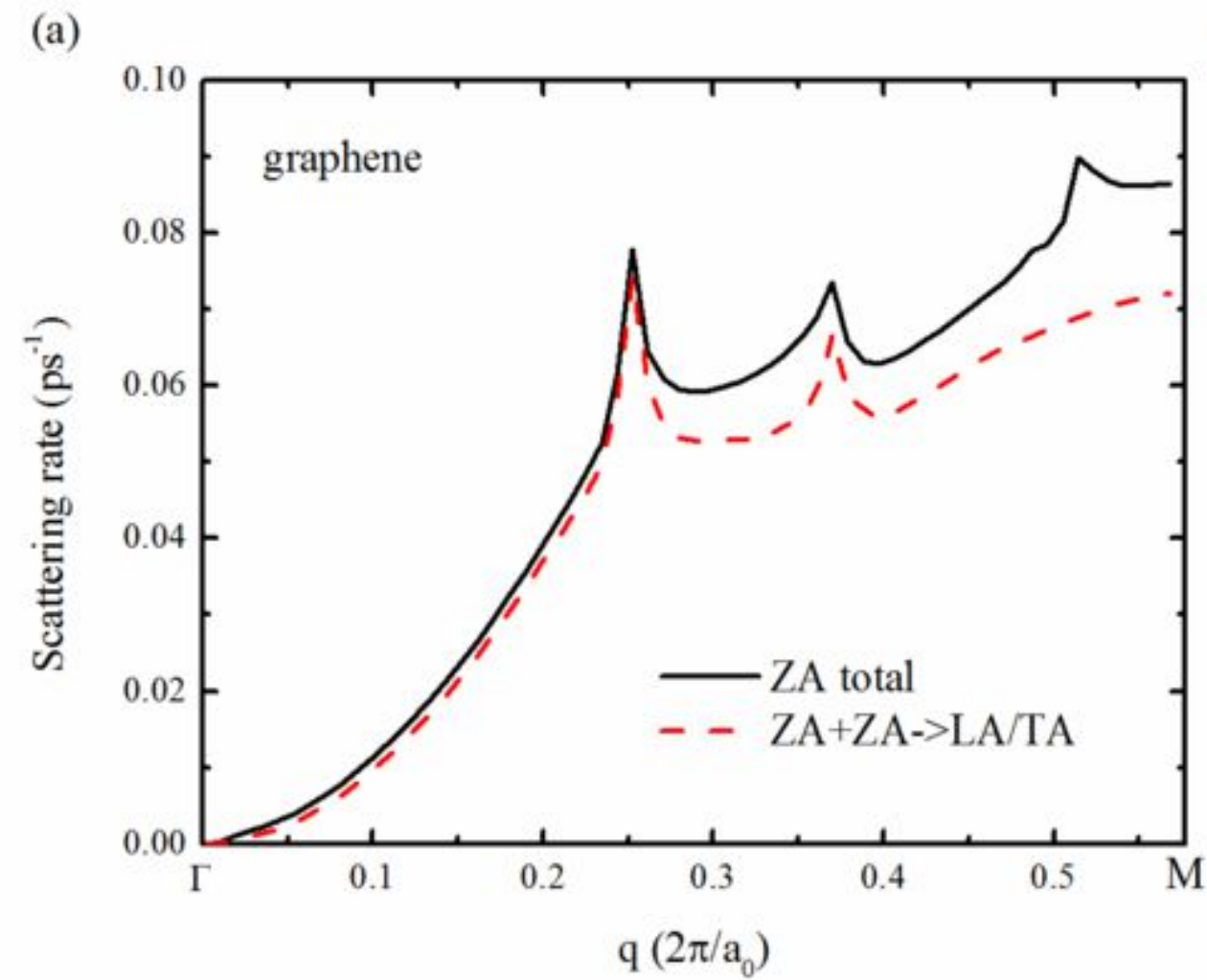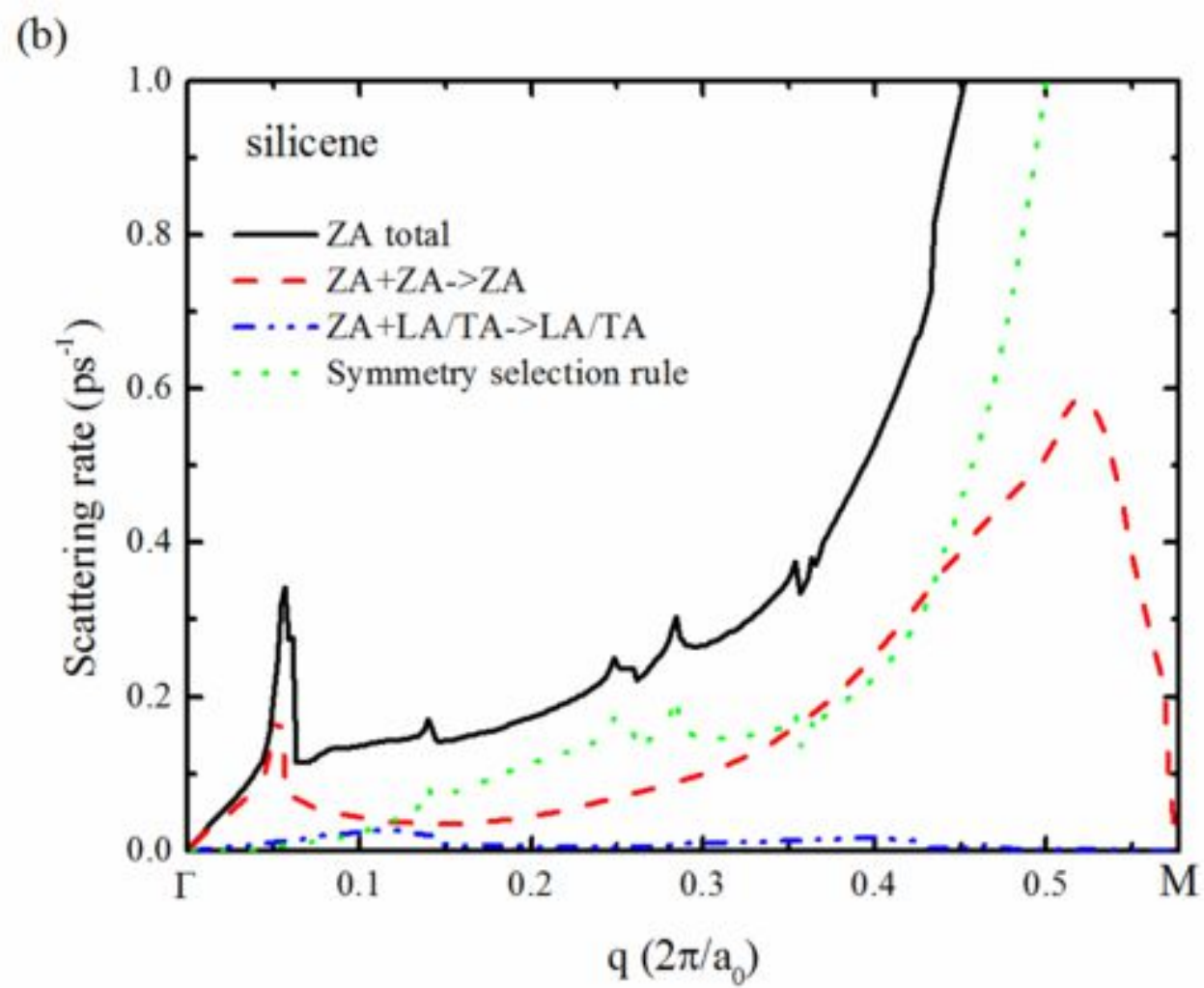

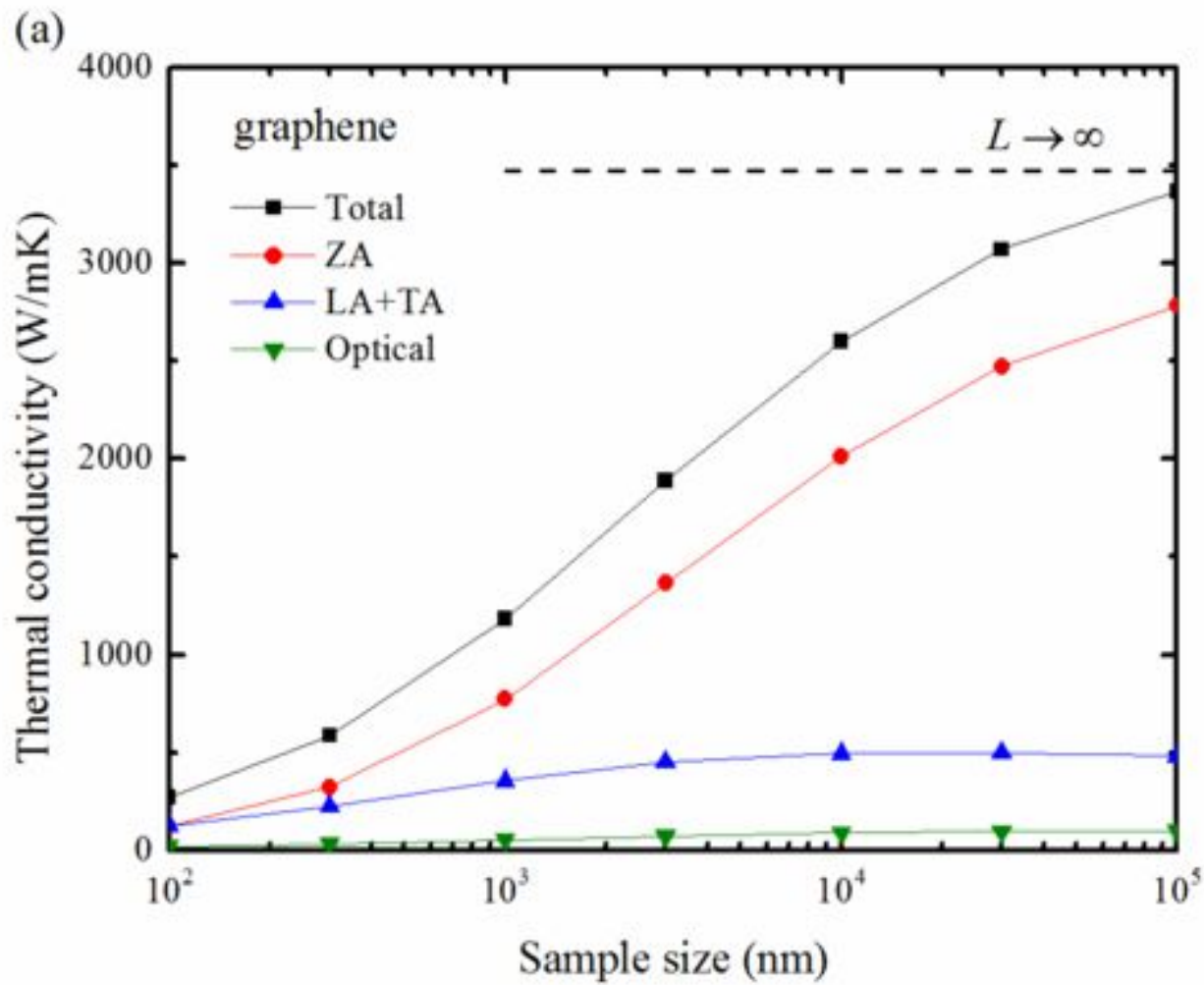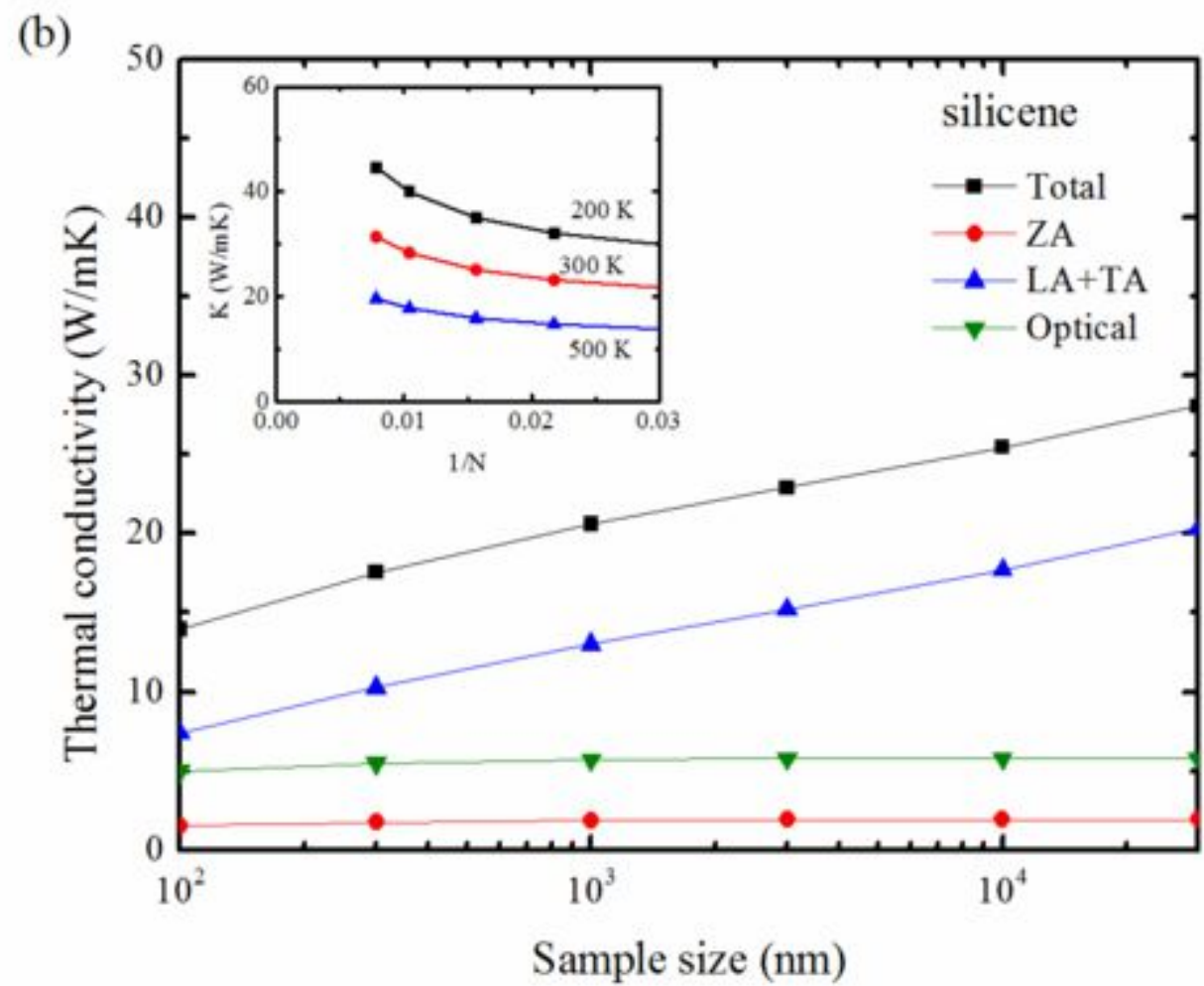

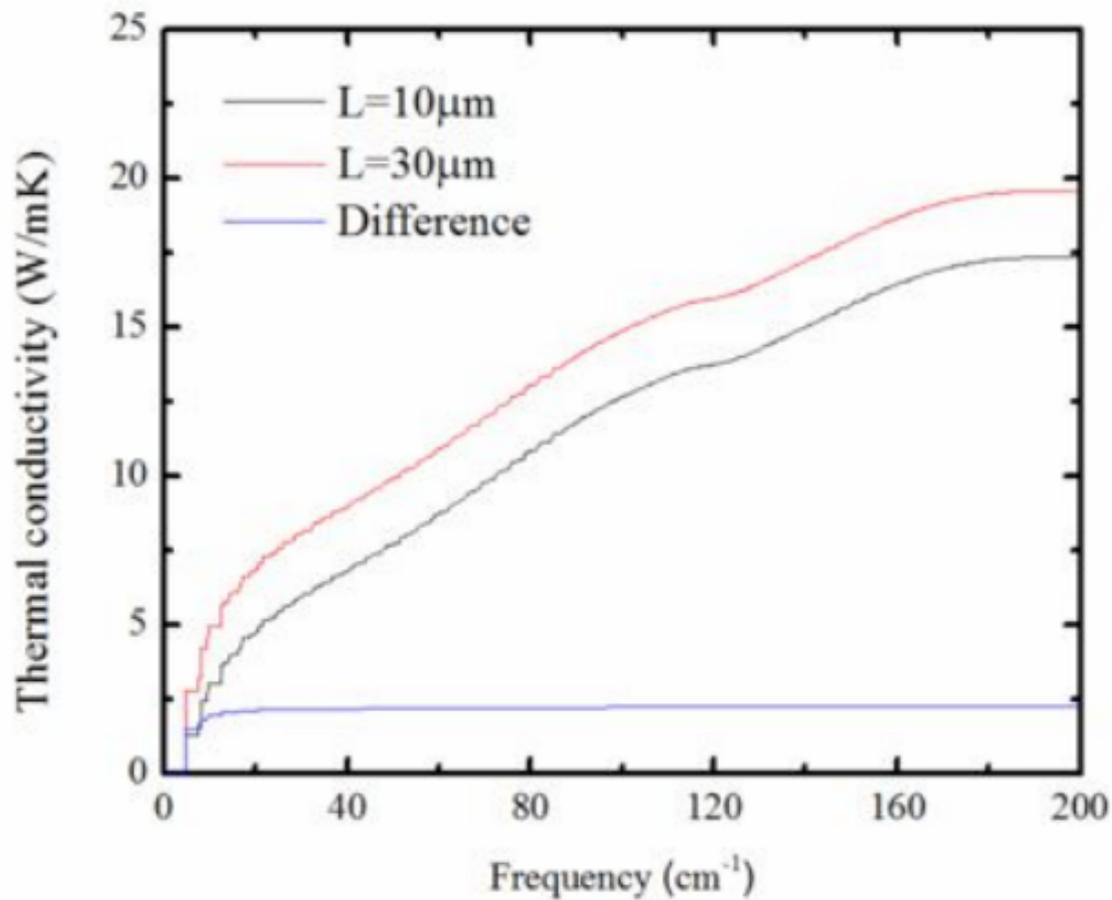

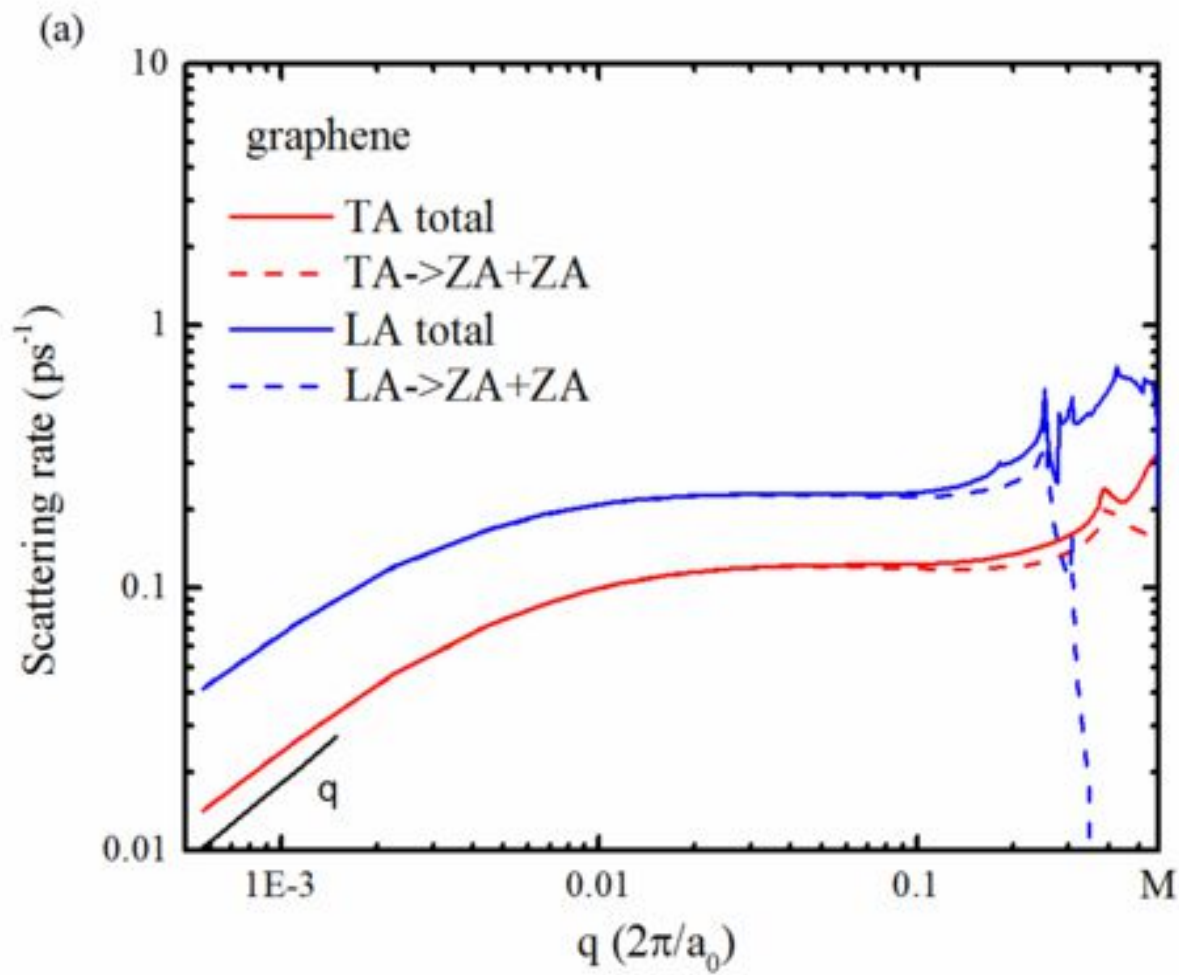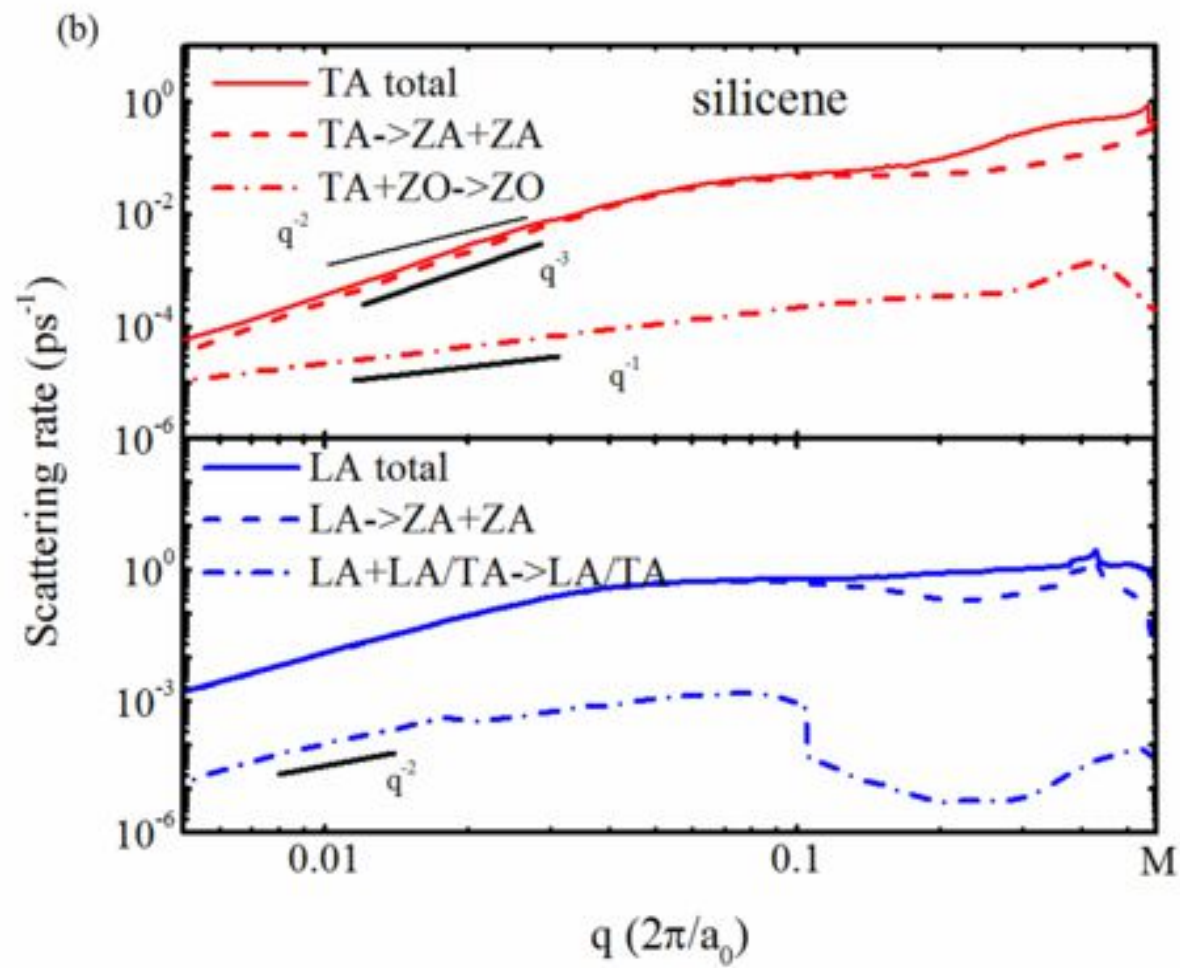

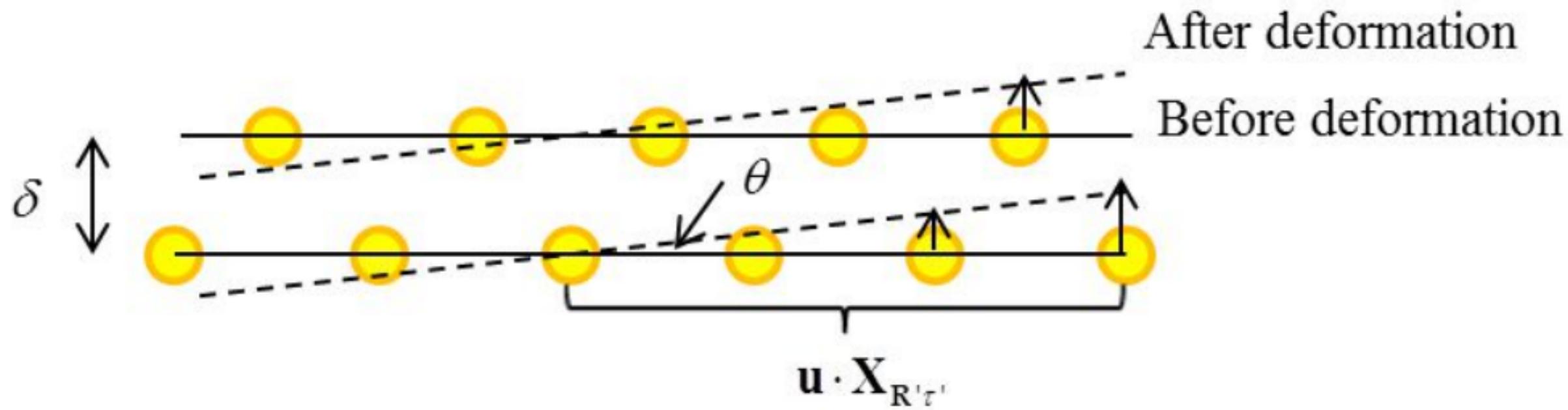